\documentclass[11pt,onecolumn]{IEEEtran}
\usepackage{epstopdf}
\usepackage{amsmath,amsfonts, amssymb,color, mathrsfs}
\usepackage{tikz, subcaption}
\usepackage[font={small,it}]{caption}
\usetikzlibrary{decorations.pathmorphing}
\usetikzlibrary{decorations.pathreplacing}
\usepackage[labelformat=simple]{subcaption}

\newtheorem{theorem}{Theorem}
\newtheorem{corollary}{Corollary}
\newtheorem{lemma}{Lemma}
\newtheorem{remark}{Remark}
\newtheorem{definition}{Definition}
\newtheorem{example}{Example}

\newtheorem{proposition}{Proposition}

\ifCLASSOPTIONcompsoc
  \usepackage[nocompress]{cite}
\else
  \usepackage{cite}
\fi
\ifCLASSINFOpdf
\else
\fi
\begin{document}
\title{\LARGE \bf The Strategic Formation of Multi-Layer Networks}

\title{\LARGE \bf The Strategic Formation of Multi-Layer Networks}
\author{Ebrahim Moradi Shahrivar~and~Shreyas Sundaram% <-this % stops a space
\IEEEcompsocitemizethanks{\IEEEcompsocthanksitem Ebrahim Moradi Shahrivar is with the Department of Electrical and Computer Engineering at the University of Waterloo. E-mail: {\tt emoradis@uwaterloo.ca}.%\protect\\
%E-mail: see http://www.michaelshell.org/contact.html
\IEEEcompsocthanksitem Shreyas Sundaram is with the School of Electrical and Computer Engineering at Purdue University.  E-mail: {\tt sundara2@purdue.edu}. Corresponding author.}% <-this % stops an unwanted space

\thanks{This material is based upon work supported by the Natural Sciences and Engineering Research Council of Canada.  A portion of the results in this paper appeared at the 52nd IEEE Conference on Decision and Control \cite{MoradiShahrivar13}.}
}

\IEEEtitleabstractindextext{%
\begin{abstract}
We study the strategic formation of multi-layer networks, where each layer represents a different type of relationship between the nodes in the network and is designed to maximize some utility that depends on the topology of that layer and those of the other layers.  We start by generalizing distance-based network formation to the two-layer setting, where edges are constructed in one layer (with fixed cost per edge) to minimize distances between nodes that are neighbors in another layer. We show that designing an optimal network in this setting is NP-hard. Despite the underlying complexity of the problem, we characterize certain properties of the optimal networks. We then formulate a multi-layer network formation game where each layer corresponds to a player that is optimally choosing its edge set in response to the edge sets of the other players.  We consider utility functions that view the different layers as strategic substitutes. By applying our results about optimal networks, we show that players with low edge costs drive players with high edge costs out of the game, and that hub-and-spoke networks that are commonly observed in transportation systems arise as Nash equilibria in this game.
\end{abstract}

}

\maketitle

\IEEEdisplaynontitleabstractindextext

% \fi
\IEEEpeerreviewmaketitle

%\IEEEraisesectionheading{\section{Introduction}\label{sec:introduction}}

%\IEEEPARstart{T}{his} demo file is intended to serve as a ``starter file''
%for IEEE Computer Society journal papers produced under \LaTeX\ using
%IEEEtran.cls version 1.8a and later.

%%%%%%%%%%%%%%%%%%%%%%%%%%%%%%%%%%%%%%%%%%%%%%%%%%%%%%%%%%%%%%%%%%%%%%%%%%%%%%%
\section{Introduction}\label{sec:introductionnew}
%\IEEEraisesectionheading{\section{Introduction}\label{sec:introductionnew}}
%\IEEEPARstart{E}{xamples} 
Examples of complex networks abound in both the natural world (e.g., ecological, social and economic systems), and in engineered applications (e.g., the Internet, the power grid, etc.). The topological structure of such networks (i.e., the relationships and interactions between the various nodes) plays a fundamental role in the functioning of the network. Early research on the structure of complex networks primarily adopted a stochastic perspective, postulating that the links between nodes are formed randomly \cite{P1,P2}. An alternative perspective, driven by the economics, computer science and engineering communities, has argued that optimization (rather than pure randomness) plays a key role in network formation.  In such settings, edges are formed {\it strategically} (either by a designer or by the nodes themselves) in order to maximize certain utility functions, resulting in networks that can be analyzed using game-theoretic notions of equilibria and efficiency \cite{P7, P8, P9, P10}.  A particularly well-studied utility function is the so-called {\it distance-based utility} introduced in \cite{P3, P4}, where the objective is to purchase edges to minimize the distances between all pairs of nodes in the network.    Subsequent works on strategic network formation  have looked at issues such as individual decision making, price of anarchy, and directed network formation  \cite{P6, P11, goyal, basar}.

While the existing literature on strategic network formation focuses predominantly on the construction of a single set of edges between the nodes, many real-world networks inherently consist of {\it multiple} layers of relationships between the same set of nodes.  Examples include friendship and professional relationships in social networks, policy influence and knowledge exchange in organizational networks \cite{P14}, and coupled communication and energy infrastructure networks \cite{P13, marzieh}.  While there has been growing research on different aspects of multi-layer networks in recent years \cite{P14, P15, P16, P17, magnani, porter}, the problem of {\it strategic multi-layer network formation} has started to receive attention only recently; aside from our initial work in~\cite{MoradiShahrivar13},  the paper \cite{wine} considers a model where each node can construct edges to a second high-speed network in order to minimize communication costs.

Here, we begin a study of strategic multi-layer network formation by generalizing distance-utility network formation to the case where one layer (or network) is formed by optimizing the distances between nodes that are neighbors in another layer (or network). As a motivating example, consider the problem in \cite{P12}, where both the physical infrastructure network and the traffic flow network between a group of cities are studied.  Interpreting traffic flow as the weight of the connection between the endpoint cities, the objective is to design an optimal infrastructure network between cities with respect to the given traffic flow pattern.  In the simplest case, this problem can be modeled as a network formation problem with a distance-based utility function where only the distances between specific pairs of nodes matter (i.e., those pairs with sufficiently high traffic flow between them). We address this class of problems by first defining a network $G_1$ capturing an existing set of relationships between nodes, and then studying the formation of an optimal second network $G_2$ based on $G_1$. We call the optimally designed network $G_2$ with respect to $G_1$ the {\it best response} network to $G_1$. 
Distance-based utilities have also been used to study social networks (where each node is an individual and the edges indicate relationships) \cite{P3, P4} and the Internet (where each node represents a router and the edges indicate communication links) \cite{P6}.  Our formulation generalizes the settings presented in those papers by allowing only distances between certain pairs of nodes (e.g., individuals in the social network or routers in the Internet) to matter when evaluating the utility of the network.   For instance, in the case of the Internet or other communication networks, the reference layer $G_1$ represents the virtual communication network indicating which pairs of nodes wish to exchange information, and the designed layer $G_2$ represents the physical communication network.

While the best response networks have been completely characterized in the case where $G_1$ is the complete network \cite{P3, P4}, we show in this paper that finding a best response network with respect to an arbitrary graph $G_1$ is NP-hard.  We characterize some useful properties of the optimal networks that arise in this setting, including upper bounds on the number of constructed edges, lower and upper bounds on the utility of the best response networks, and conditions for the empty network to be a best response.   These properties enable us to find best response networks with respect to certain specific reference networks, i.e., forests and networks with a star subgraph.   

We then use the notion of the best response network to model a scenario with multiple network designers, each of whom is building a different layer of the network. An example of this is when multiple transportation companies build their individual service networks among a group of cities, and each company prefers to provide service between pairs of cities that are not already covered by other companies.  We capture these scenarios by defining a non-cooperative multi-layer network formation game where each player corresponds to a specific layer of the network.  We develop a notion of distance-based multi-layer network formation based on {\it strategic substitutes}, where the presence of an edge in one layer makes it less desirable to have that edge in another layer.  Despite the complexity of calculating best response networks, we characterize the Nash equilibrium networks that arise in this setting. In particular, we show that players with low costs for building edges drive out players that have relatively high costs, and that our framework gives rise to the ``hub-and-spoke'' networks commonly seen in various transportation systems \cite{jeanpaul}.

\section{Definitions}
\label{sec:defs}
An undirected network (or graph) is denoted by $G= (N, E)$ where $N=\{ v_1, v_2, \dots , v_n\}$ is the set of nodes (or vertices) and $E \subseteq \{(v_i,v_j) \lvert v_i,v_j \in N, v_i \neq v_j \}$. The set of all possible graphs on $N$ is denoted by $G^N$. Two nodes are said to be neighbors if there is an edge between them. The degree of a node $v_i \in N$ is the number of its neighbors in graph $G$, and is denoted by $\deg_i(G)$. A leaf node is a node that has degree one, i.e., it has only one neighbor. A path from node $v_1$ to $v_k$ in graph $G$ is a sequence of distinct nodes $v_1 v_2 \cdots v_k$ where there is an edge between each pair of consecutive nodes of the sequence. The length of a path is the number of edges in the sequence. We denote the shortest distance between nodes $v_i$ and $v_j$ in graph $G$ by $d_G (i,j)$. If there is no path from $v_i$ to $v_j$, we take $d_G(i,j)=\infty$. The diameter of the graph $G$ is $\max_{v_i,v_j \in N, v_i \ne v_j}d_G(i,j)$.  A cycle is a path of length two or more from a node to itself. A graph $G'= (N', E')$ is called a subgraph of $G=(N, E)$, denoted as $G' \subseteq G$, if $N' \subseteq N$ and $E' \subseteq E \cap \{ N' \times N' \}$. A graph $G'$ is said to be induced by a set of nodes $N' \subseteq N$ if $E'=E \cap \{ N' \times N' \}$.  A graph is connected if there is a path from every node to every other node.   A subgraph $G' = (N', E')$ of $G$ is a component if $G'$ is connected and there are no edges in $G$ between nodes in $N'$ and nodes in $N \setminus N'$.

A tree is a connected acyclic graph. For a connected graph $G=(N, E)$, a connected acyclic subgraph $T = (N, E_T)$ of $G$ is called a spanning tree of $G$.  
A spanning forest of a disconnected graph is a collection of spanning trees of each of its components.

We denote the complete graph (i.e., the graph with an edge between every pair of different nodes) by $G^c=(N, E^c)$. We use $G^e =(N, \phi)$ to denote the empty graph. Finally, $G^s=(N, E^s)$ is a star graph, which is a tree graph with one node that is connected to all other nodes. The complement of graph $G = (N,E)$ is denoted by $\sim{G} = (N,\sim{E})$, where $ \sim{E} \triangleq E^c \setminus E$. Two graphs on the same set of nodes are said to be disjoint if their edge sets are disjoint.

\section{Distance-Based Utility}
\label{sec:distance_utility}
A canonical problem in network formation introduced by Jackson and Wolinsky involves distance-based utilities \cite{P3}. In this model, there is a net benefit of $b(k)$ for each pair of nodes that are $k$ hops away from each other in the network, where $b:\{1,2,\cdots,n-1,\infty\} \to \mathbb{R}_{\geq 0}$ is a real-valued nonincreasing nonnegative function (i.e., nodes that are further away from each other provide smaller benefits) and $b(\infty)=0$. There is a cost $c \in \mathbb{R}_{> 0}$ for each edge in the network. The outcome of the network formation process is a graph $G=(N, E) \in G^N$.  The utility (or value) of a given graph $G \in G^N$ is given by the utility function 
\begin{equation}\label{Prob1}
u(G)=\sum_{ \substack{v_i,v_j \in N\\
					     v_i \ne v_j}} b(d_G(i,j))- c |E|.
\end{equation}
In this formulation, there is an inherent trade-off faced by the designer: adding links to a larger number of nodes provides a larger benefit (by reducing the distances between nodes), but also incurs a larger cost invested in links. An {\it optimal} (or efficient) network $G$ satisfies $u(G) \geq u(G')$, $\forall G' \in G^N$.

The following result from \cite{P3, P4} shows that when $b(\cdot)$ is a strictly decreasing function, there are only a few different kinds of efficient networks, depending on the relative values of the link costs and  connection benefits. 
\begin{proposition}\cite{P4}\label{prop:jacksonresult}
Assume that $b(\cdot)$ is a strictly decreasing function. In the distance-based utility model,
\begin{itemize}
\item if $c<b(1)-b(2)$, then the complete network is the unique efficient network;
\item if $b(1)-b(2)<c<b(1)+(n-2)b(2)/2$, then the star is the unique efficient network;
\item if $b(1)+(n-2)b(2)/2<c$, then the empty network is the unique efficient network.
\end{itemize}
\end{proposition}
In the above proposition, whenever $c$ is equal to one of the specified upper or lower   bounds, there will be more than one efficient network: if $c=b(1)-b(2)$, then the complete network and star network are both efficient, and if $c=b(1)+(n-2)b(2)/2$, the star network and the empty network are both efficient networks with zero utility.  Furthermore, for the more general case where $b(\cdot)$ is nonincreasing, the three networks given by the above result are still optimal for the corresponding ranges of costs and benefits, although they may no longer be unique.

In the next sections, we will generalize the distance utility framework to the two-layer network formation setting.  We will characterize the complexity of determining efficient networks in such settings and provide properties of such networks.  We will then apply these results to study a multi-layer network formation game with multiple network designers.

%%%%%%%%%%%%%%%%%%%%%%%%%%%%%%%%%%%%%%%%%%%%%%%
\section{Two Layer Distance-Based Utilities:  \\Best Response Network}
\label{sec:two_layer}
In the traditional distance-based network formation problem described above, the objective is to minimize the distances between every possible pair of nodes.  However, in many settings, one is only interested in minimizing distances between {\it certain} pairs of nodes.  For example, consider a communications system where each node only wishes to exchange information with a subset of the other nodes, and the task is to design a physical network to provide short paths between those pairs of nodes. To handle these types of scenarios, in this section we generalize the study of distance-based network formation to a {\it multi-layer} setting.  Specifically, suppose that we have a layer (or graph) $G_1=(N, E_1)$, where the edge set $E_1$ specifies a type of relationship between the nodes in $N$.  Our objective is to design another layer (or graph) $G = (N,E)$ on the same set of nodes, where the utility of the graph is given by
\begin{equation}\label{P_2}
u(G|G_1)=\sum_{(v_i,v_j) \in E_1} b(d_G(i,j))- c |E|.
\end{equation}
Note that the summation is only over edges in set $E_1$, capturing the fact that only distances between those pairs of nodes matter in graph $G$;  the traditional distance utility function in \eqref{Prob1} is obtained as a special case when $G_1$ is the complete graph.  

Assume $G_2 = (N, E_2)$ is a network that maximizes \eqref{P_2}; we say $G_2$ is a {\it best response} (BR) network to $G_1$, or equivalently, an {\it efficient} network with respect to the utility function \eqref{P_2}.

\begin{remark}
The utility function~\eqref{P_2} does not necessarily have a unique maximizer; indeed, in many cases, there are multiple best response networks with respect to a given network, as demonstrated by Example~\ref{example1} below.
\end{remark}

When $G_1$ is the complete network, the best response is trivially a subgraph of $G_1$.  However, the following example demonstrates that the best response network to a general network $G_1$ does not necessarily have to be a subgraph of that network.

\begin{figure}[hb]
\begin{minipage}[b]{.33\textwidth}
\begin{center}
\begin{tikzpicture}
  [scale=.3, inner sep=1pt, minimum size=3pt, auto=center,every node/.style={circle, draw=black, thick}]
  \node (n1) at (6,7)  {\footnotesize $v_1$};
  \node (n2) at (3,6)  {\footnotesize $v_2$};
  \node (n6) at (9,6)  {\footnotesize $v_6$};
  \node (n3) at (3,2)  {\footnotesize $v_3$};
  \node (n5) at (9,2)  {\footnotesize $v_5$};
  \node (n4) at (6,1)  {\footnotesize $v_4$};

  \foreach \from/\to in {n1/n2,n2/n3,n3/n4,n4/n5,n5/n6,n6/n1}
    \draw (\from) -- (\to);
\end{tikzpicture}
\end{center}
     \subcaption{$G_1$}
     \label{fig:best_response_not_subgraph_a}
\end{minipage}%
\begin{minipage}[b]{0.33\textwidth}
\begin{center}
\begin{tikzpicture}
  [scale=.3, inner sep=1pt, minimum size=3pt, auto=center,every node/.style={circle, draw=black, thick}]
  \node (n1) at (6,7)  {\footnotesize $v_1$};
  \node (n2) at (3,6)  {\footnotesize $v_2$};
  \node (n6) at (9,6)  {\footnotesize $v_6$};
  \node (n3) at (3,2)  {\footnotesize $v_3$};
  \node (n5) at (9,2)  {\footnotesize $v_5$};
  \node (n4) at (6,1)  {\footnotesize $v_4$};

  \foreach \from/\to in {n1/n2,n1/n3,n1/n4,n1/n5,n6/n1}
    \draw (\from) -- (\to);
\end{tikzpicture}
\end{center}
     \subcaption{$G_2$}
 \label{fig:best_response_not_subgraph_b}
\end{minipage}
\begin{minipage}[b]{.33\textwidth}
\begin{center}
\begin{tikzpicture}
  [scale=.3, inner sep=1pt, minimum size=3pt, auto=center,every node/.style={circle, draw=black, thick}]
  \node (n1) at (6,7)  {\footnotesize $v_1$};
  \node (n2) at (3,6)  {\footnotesize $v_2$};
  \node (n6) at (9,6)  {\footnotesize $v_6$};
  \node (n3) at (3,2)  {\footnotesize $v_3$};
  \node (n5) at (9,2)  {\footnotesize $v_5$};
  \node (n4) at (6,1)  {\footnotesize $v_4$};

  \foreach \from/\to in {n1/n2,n4/n3,n1/n4,n4/n5,n6/n1}
    \draw (\from) -- (\to);
\end{tikzpicture}
\end{center}
 \subcaption{$G_3$}
     \label{fig:best_response_not_subgraph_c}
\end{minipage}

\caption{Illustration of potential best response networks with respect to network $G_1$.}
\end{figure}
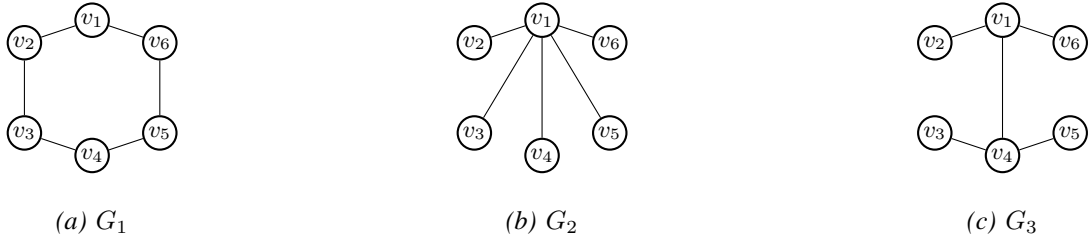

\begin{example}\label{example1}
Consider the ring graph $G_1$ with 6 nodes shown in Figure~\ref{fig:best_response_not_subgraph_a}.   Suppose $b(1)=c + \epsilon$, for some small constant $\epsilon > 0$.  Then,
\begin{enumerate}
\item The utility~\eqref{P_2} of $G_1$ to itself is $u(G_1 | G_1)= 6(b(1)-c)= 6 \epsilon$. 
\item Any subgraph of $G_1$ with 5 edges is a path graph.  This has utility $5(b(1)-c) + b(5) = 5\epsilon + b(5)$.
\item Any subgraph of $G_1$ with $k$ edges, where $k<5$, has utility $k(b(1)-c) = k \epsilon$.
\end{enumerate}
Thus, when $b(5) > \epsilon$, the best subgraph of $G_1$ is the path graph with the utility given above.

Now, the star graph shown in Figure~\ref{fig:best_response_not_subgraph_b} has utility $2b(1)+4b(2)-5c$. This is better than the path graph if $4b(2)-3b(1)>b(5)$, which holds, for example, when $b(2)$ is sufficiently close to $b(1)$ and $b(2) > b(5)$.  Therefore, for  utility functions that satisfy this property, no subgraph of $G_1$ can be a BR to $G_1$.

For certain benefit functions a star is not a BR either.  The graph $G_3$ given in Figure~\ref{fig:best_response_not_subgraph_c} has utility $4b(1)+2b(3)-5c$.  This is better than the path graph if $2b(3)-b(1)>b(5)$, and better than the star if $b(3)>2b(2)-b(1)$. For instance if $c=1$,  $b(1)=1.01, b(2)=0.85, b(3)=0.8, b(4)=0.2$ and $b(5)=0.1$, then the graph $G_3$ is better than the star graph or any subgraph of $G_1$, i.e., $u(G_3|G_1)>u(G|G_1)$ where $G \subseteq G_1$ or $G=G_2$. In this example, one can verify (e.g., using a brute-force search) that $G_3$ is in fact a BR network to $G_1$.

It is also instructive to consider the case where $b(1)=b(2)=b(3)>\operatorname*{max} \{c, b(4)\}$. In this case, the graphs shown in Figure~\ref{fig:best_response_not_subgraph_b} and \ref{fig:best_response_not_subgraph_c} are both best response networks to $G_1$ and have higher utility than any subgraph of $G_1$.
\end{example}

The above example illustrates that BR networks to an arbitrary graph $G_1$ are very sensitive to the relative values of the benefit function $b(\cdot)$ and the cost $c$. Indeed, the shape of the entire benefit function can play a role in determining the best response to general graphs, whereas only the value of $b(1)$ and $b(2)$ matter when $G_1$ is the complete graph (as shown in Proposition \ref{prop:jacksonresult}). One of the main results of this paper is to formally characterize the complexity of finding a best response network to a given graph. To do this, we first cast it as a decision problem (i.e., a question to which the answer is {\it yes} or {\it no}) as follows. 

\begin{definition}\label{BR}
{\bf Best Response Network (\textbf{BRN}) Problem.} \\
INSTANCE: A network $G_1=(N, E_1)$, a nonincreasing benefit function $b:\{1,2,\cdots,n-1,\infty\} \to \mathbb{R}_{\geq 0}$, an edge cost $c \in \mathbb{R}_{> 0}$ and a lower bound on utility given by $r \in \mathbb{R}_{>0}$.\\
QUESTION: For the utility function $u(\cdot)$ given in equation \eqref{P_2}, does there exist a $G=(N, E) \in G^N$ such that
\begin{align} \label{BRNeq}
u(G|G_1) \geq r?
\end{align}
\end{definition}

Assuming that the input size to a problem is $n$, if there is an algorithm that solves the problem in $O(n^k)$ time (for some positive constant $k$), the problem is said to be in the complexity class P. A decision problem is said to be in the class NP if every ``yes" answer has an accompanying certificate that can be verified in polynomial-time. Consider two decision problems $A$ and $B$ and assume that there exists a polynomial-time transformation from any instance $b$ of problem $B$ into some instance $a$ of problem $A$ such that the answer to $b$  is ``yes" if and only if answer to $a$ is ``yes". If such a transformation from $B$ to $A$ exists, it is called a {\it reduction} and problem $B$ is said to be {\it polynomial-time reducible} to problem $A$. A problem $A$ is NP-hard if for all problems $B \in$ NP, $B$ is polynomial-time reducible to $A$; in particular, $A$ is NP-hard if some other NP-hard problem $B$ is polynomial-time reducible to $A$ \cite{CLRS}. An NP-hard problem that is also in the class NP is said to be NP-complete. The following theorem is one of our main results and shows that finding a BR with respect to an arbitrary graph with arbitrary cost and nonincreasing benefit functions does not have a polynomial-time solution, unless the answer to the long-standing open question of whether P = NP is affirmative.

\begin{theorem}\label{complexityBR}
BRN is NP-hard.
\end{theorem}

We will develop the proof of Theorem~\ref{complexityBR} over the rest of this section. We will require some intermediate properties of best response networks, given by the following results.

%%%%%%%%%%%%%%%%%%%%%%%%%%%%%%%%%%%%%%%%%%%%%%%%%%%%%%%%%%%%%%%%%

\subsection{Some Properties of Best Response Networks}

\begin{lemma} \label{lem1}
If $G_2=(N, E_2)$ is a BR network to $G_1=(N, E_1)$, then the number of edges in $G_2$ is less than or equal to the number of edges in $G_1$. If $b(1)>b(2)$, then $G_1$ and $G_2$ have an equal number of edges if and only if $G_2 = G_1$.
\end{lemma}
\begin{IEEEproof}
We use contradiction to prove the first part. Suppose that $G_2$ is a BR and has more edges than $G_1$. Then
\begin{align*}
u(G_2|G_1) &= \sum_{(u,v) \in E_1} b(d_{G_2}(u,v)) - c|E_2| \\
&\leq |E_1|b(1)-c|E_2|\\
&< |E_1|b(1)-c|E_1|= u(G_1|G_1),
\end{align*}
which contradicts our assumption that $G_2$ is a BR to $G_1$.
To prove the second part, note that if $G_2=G_1$ then the number of edges in $G_2$ and $G_1$ are equal. So we only need to show that when $b(1)>b(2)$, if the number of edges in $G_2$ is equal to the number of edges in $G_1$, then $G_2=G_1$. If $G_2 \neq G_1$, then there exists a $(u,v) \in E_1$ such that $d_{G_2} (u,v) \geq 2$. Thus
\begin{align*}
 u(G_2|G_1) &= \sum_{(u,v) \in E_1} b(d_{G_2}(u,v)) - c|E_2| \\
&< |E_1|b(1)-c|E_1|=u(G_1|G_1),
\end{align*}
contradicting the assumption that $G_2$ is a BR to $G_1$. 
\end{IEEEproof}

The next lemma discusses the connectivity of BR networks.

\begin{lemma}\label{connectedgraph}
Suppose that $G_2$ is a best response network to $G_1$ and $b(1)>c$. Then any two nodes that are connected by a path in $G_1$ will also be connected by a path in $G_2$. Specifically, if $G_1$ is connected, then $G_2$ must be connected.
\end{lemma}
\begin{IEEEproof}
Let $u$ and $v$ be two nodes that are neighbors in $G_1$. By way of contradiction assume that there is no path between $u$ and $v$ in the BR network $G_2=(N, E_2)$. For $G'_2=(N, E'_2)$ with $E'_2=E_2 \cup \{(u,v)\}$,
\begin{equation*}
u(G'_2|G_1)-u(G_2|G_1) \geq b(1)-c>0,
\end{equation*}
contradicting the assumption that $G_2$ is a BR network. Now consider the case that $u$ and $v$ are connected through a path in $G_1$. Then there must be a path from $u$ to $v$ in $G_2$, since we showed that any two nodes that are directly connected in $G_1$ remain connected in $G_2$.
\end{IEEEproof}

\begin{remark}\label{equalbc}
When $b(1) = c$, the above proof can be applied to show that there exists a best response network in which any two nodes that are connected by a path in $G_1$ will also be connected by a path in $G_2$ (although this does not have to be true of {\it every} best response network).
\end{remark}

For any integer $t \geq 1$, a subgraph $H=(N, E_H)$ of $G_1=(N, E_1)$ is called a {\it $t$-spanner} if $d_H(x,y) \leq t$ for all $(x,y) \in E_1$, i.e., the distance between each pair of nodes that are neighbors in $G_1$ is not more than $t$ in $H$ \cite{Cai}. A subgraph $T=(N, E_T)$ of the graph $G_1$ that is both a $t$-spanner and a tree is called a {\it tree $t$-spanner}.  The following important lemma characterizes a BR to graphs that have a 2-spanner.

\begin{lemma}\label{lem:spanforestresult}
Suppose graph $G_1=(N, E_1)$ has a spanning forest\footnote{Whenever $G_1$ is a connected network, by a spanning forest of $G_1$ we mean a spanning tree.} $F=(N, E_F)$  that is also a 2-spanner. Assume that $b(1)-b(2)\leq c \leq b(1)$. Then $F$ is a BR to $G_1$.
\end{lemma}
\begin{IEEEproof}
Assume that $G_1$ has $m$ components where $m \geq 1$. Since $F$ is a spanning forest, $|E_F|=|N|-m$. Using the fact that $d_F(x,y) \leq 2$ for all $(x,y) \in E_1$, we have
\begin{equation}\label{spanforest}
u(F|G_1)=(|N|-m)(b(1)-c)+(|E_1|-(|N|-m))b(2).
\end{equation}
Now assume that $H=(N, E_H)$ is a best response network to $G_1$ such that any two nodes that are connected in $G_1$ are also connected in $H$. The existence of such a BR network is guaranteed by Lemma~\ref{connectedgraph} and Remark~\ref{equalbc}. Thus $|E_H| \geq |N|-m$. Also by Lemma \ref{lem1}, we have $|E_H| \leq |E_1|$. Since at most $|E_H|$ pairs of neighbors in $G_1$ can be directly connected in $H$, the remaining $|E_1|-|E_H|$ pairs of neighbors in $G_1$ will be at least a distance of two away from each other in $H$.  Thus we have 
\begin{align}\label{eqlimit}
u(H|G_1)&\leq |E_H|(b(1)-c)+(|E_1|-|E_H|)b(2) \\
&= (|N|-m)(b(1)-c)+(|E_H|-(|N|-m))(b(1)-c)+(|E_1|-|E_H|)b(2)\nonumber \\
&\leq (|N|-m)(b(1)-c)+(|E_1|-(|N|-m))b(2)\nonumber \\
&= u(F|G_1).\nonumber 
\end{align}
Thus $F$ is a BR to the network $G_1$.
\end{IEEEproof}

The next lemma provides lower and upper bounds on the utility of BR networks when $b(1)-b(2)\leq c \leq b(1)$. 

\begin{lemma}\label{limits}
Suppose that $b(1)-b(2)\leq c \leq b(1)$ and $G_2=(N, E_2)$ is a BR network with respect to an arbitrary connected network $G_1=(N, E_1)$. Then
\begin{align} 
|E_1|(b(1)-c)  \leq u(G_2|G_1) \leq (|N|-1)(b(1)-c)  
+(|E_1|-|N|+1)b(2). \label{newinq} 
\end{align}
\end{lemma}
\begin{IEEEproof}
The lower bound follows from the fact that $u(G_2|G_1) \ge u(G_1|G_1) = |E_1|(b(1)-c)$, by virtue of $G_2$ being a BR network.  For the upper bound, note that since $b(1) \geq c$, $G_2$ can be assumed to be a connected graph (by Lemma~\ref{connectedgraph} and Remark \ref{equalbc}) and thus $|E_2| \geq |N|-1$. The rest of the proof follows the same procedure as in the proof of Lemma~\ref{lem:spanforestresult} with $m=1$.
\end{IEEEproof}

\begin{remark}
The inequalities given in the above lemma are sharp.  As we will show later in this paper, a BR to a tree is the same tree if $b(1) \geq c$. For a tree, the left and right hand sides of inequality \eqref{newinq} are equal. Also, for a graph $G_1$ with a tree  2-spanner $T$, we know that $T$ is a BR to $G_1$ by Lemma~\ref{lem:spanforestresult} with utility equal to the right hand side of inequality \eqref{newinq}.
\end{remark}

%%%%%%%%%%%%%%%%%%%%%%%%%%%%%%%%%%%%%%%%%%%%%%%%%%%%%%%%

\subsection{Proof of NP-Hardness of the BRN Problem}

We now return to the BRN problem (Definition~\ref{BR}) and the claim of NP-hardness given in Theorem~\ref{complexityBR}.  To prove this theorem, we will construct a reduction from the {\it Tree $t$-spanner Problem} \cite{Cai}, defined below.

\begin{definition}
{\bf Tree $t$-Spanner (\textbf{TtS}) Problem.} \\
INSTANCE: A connected graph $G=(N, E)$ and a positive integer $t$.\\
QUESTION: Does $G$ have a tree $t$-spanner, i.e., a subgraph $T=(N, E_T)$ such that $|E_T|=|N|-1$ and $d_T(x,y) \leq t$ for all $(x,y) \in E$?
\end{definition}

The TtS problem is in P for $t=2$, but NP-complete for all $t \geq 4$; the complexity of the problem for $t=3$ is still unknown \cite{Cai}. We are now in place to prove Theorem \ref{BR}. 

\begin{IEEEproof}[Proof of Theorem 1]
We will construct a reduction from the TtS problem to the BRN problem, which will then imply that the BRN problem is NP-hard. Consider an instance of the TtS problem with graph $G=(N, E)$ and $t=4$. Any spanning tree of $G$ with $|N| \leq 5$ is a tree 4-spanner which is easy to find. Thus, we assume that $|N| \geq 6$. Define the corresponding instance of the BRN problem as follows. The network $G_1=(N, E_1)$ is the same as the graph $G$, i.e., $G_1=G$. The benefit function $b(\cdot)$ and edge-cost $c$ are chosen to satisfy
\begin{equation}
\begin{split}
b(1)>b(2)=b(3)&=b(4)>b(5),\\
b(1) - b(2)< c& <b(1).
\end{split}
\label{BRNinst}
\end{equation}
For example $c = 2$, $b(1)=3$, $b(2)=b(3)=b(4)=2$ and $b(k)=0~\forall k \geq 5$ satisfies these conditions. Finally set 
\begin{equation}\label{threshold}
r=(|N|-1)(b(1)-c)+(|E_1|-(|N|-1))b(2).
\end{equation}
Clearly we can construct the above BRN instance in polynomial time. Now assume that the answer to the instance of the TtS problem is ``yes", i.e., graph $G$ has a tree 4-spanner $T=(N, E_T)$. This means that $T$ is a subtree of $G_1$ and $d_T(x,y) \leq 4$ for all $(x,y) \in E_1$. Thus we have that
\begin{align*}
u(T|G_1)&=\sum_{(x,y)\in E_1\setminus E_T} b(d_T(x,y))+(|N|-1)(b(1)-c)\\
&=(|E_1|-(|N|-1))b(2)+(|N|-1)(b(1)-c)\\
&=r.
\end{align*}
Note that we used the fact that $b(2)=b(3)=b(4)$ to go from the first line to the second line in the above equation. Therefore, the answer to the defined instance of the BRN problem is also ``yes''. 

To complete the proof, we have to show that if the answer to the constructed instance of the BRN is ``yes'', then the answer to the instance of the TtS is ``yes''. In other words, we have to show that if there exists a graph $G_2=(N, E_2)$ such that
\begin{equation*}
u(G_2|G_1)=\sum_{(x,y)\in E_1} b(d_{G_2}(x,y)) -c|E_2| \geq r,
\end{equation*}
where $b(\cdot)$ and $c$ satisfy \eqref{BRNinst} and $r$ is given by \eqref{threshold}, then $G_1$ has a tree 4-spanner. We claim that any $G_2$ with utility at least $r$ must be a tree 4-spanner of $G_1$.

Assume that $G_2=(N, E_2)$ is a graph with $u(G_2|G_1)\geq r$. Since $r$ is equal to the upper bound of the utility of the BR (by Lemma \ref{limits}), $G_2$ must be a best response to $G_1$.  Since $b(1)>c$, by Lemma \ref{connectedgraph} we know that $G_2$ is a connected graph. Therefore, $|E_2| \geq |N|-1$. First consider the case that $|E_2|>|N|-1$. Then similar to equation \eqref{eqlimit}, we have that
\begin{align*}
u(G_2|G_1)&\leq |E_2|(b(1)-c)+(|E_1|-|E_2|)b(2)\\
&=(|N|-1)(b(1)-c)+(|E_2|-(|N|-1))(b(1)-c)+(|E_1|-|E_2|)b(2)\\
&<(|N|-1)(b(1)-c)+(|E_2|-(|N|-1))b(2)+(|E_1|-|E_2|)b(2)\\
&=(|N|-1)(b(1)-c)+(|E_1|-(|N|-1))b(2)=r,
\end{align*}
which is a contradiction. Thus consider the case that $|E_2|=|N|-1$, i.e., $G_2$ is tree. Denoting $|E_2 \cap E_1|=\gamma$, we have
\begin{align}\label{eqlimit2}
u(G_2|G_1)&=\gamma (b(1)-c)-(|N|-1-\gamma)c+\sum_{(x,y) \in E_1 \setminus E_2} b(d_{G_2}(x,y)) \\
&\leq \gamma (b(1)-c)-(|N|-1-\gamma)c+(|E_1|-\gamma)b(2)\nonumber \\
&= \gamma (b(1)-c)+(|N|-1-\gamma)(b(2)-c)+(|E_1|-(|N|-1))b(2).\nonumber 
\end{align}
If $\gamma<|N|-1$, since $b(1)-c>b(2)-c$, by equation \eqref{eqlimit2} we have that
\begin{align*}
u(G_2|G_1) &< \gamma (b(1)-c)+(|N|-1-\gamma)(b(1)-c)+(|E_1|-(|N|-1))b(2)=r,
\end{align*}
which is again a contradiction. Therefore,  $|E_2 \cap E_1|=\gamma=|N|-1$. This means that $G_2$ is a subtree of $G_1$. Now if there exists $(u,v)\in E_1$ such that $d_{G_2}(u,v)>4$, then we have 
\begin{align*}
u(G_2|G_1) &=(|N|-1) (b(1)-c)+\sum_{(x,y) \in E_1 \setminus E_2} b(d_{G_2}(x,y)) \\
&< (|N|-1) (b(1)-c)+(|E_1|-(|N|-1))b(2)\\
&=r,
\end{align*}
where the last inequality follows from the fact that $b(2)=b(3)=b(4)>b(d)$ for all $d>4$.
Therefore, for all $(u,v) \in E_1$, $d_{G_2}(u,v) \leq 4$ which means that $G_2$ must be a tree 4-spanner for the graph $G_1$.
Thus the answer to the instance of the TtS problem is ``yes". This shows that the NP-hard problem TtS (for $t = 4$) is polynomial-time reducible to BRN, and therefore BRN is  NP-hard.
\end{IEEEproof}

There are certain NP-hard optimization problems (e.g., minimum vertex cover) whose solutions can be approximated to within a constant factor by simple greedy algorithms \cite{CLRS}. The following example considers a natural greedy algorithm where edges are added or removed one at a time, and shows that this algorithm can produce results that are arbitrarily far away from the optimal network.

\begin{example}\label{greedy}
Consider a greedy algorithm where at each step, we add or remove a link that provides the highest increase in the utility until no further improvements can be made.  The following scenarios illustrate the pitfalls of such an algorithm.

Consider a reference network $G_1$.  Suppose we attempt to build a BR network by starting with an empty network $G$ and repeatedly adding edges.  If $b(1) < c$, then adding any single edge to $G$ will result in negative utility, and thus the algorithm stops with the empty network.  Since there can exist nonempty BR networks when $b(1) < c$ whose utility is unbounded in $n$ (e.g., see Proposition~\ref{prop:jacksonresult}), the network produced by the above algorithm can be arbitrarily bad in comparison to the true BR network.

Now suppose that we attempt to build a BR network by starting with the reference network $G_1$ and removing edges one at a time.  Consider the graph $G_1$ depicted in Figure~\ref{fig:best_a} and define $c = 1$, $b(1) = \frac{n-1}{n-2}$, $b(2) = 0.5$, $b(k) = 0$ for $3 \le k \le n-1$.

Starting with $G_1$, removing any of the edges increases the utility by $b(2)-(b(1)-c)$. Thus any edge is a candidate for removal. Consider removing the edge $(v_1,v_2)$ which results in network $G_2$. Now no further improvements are possible by adding or removing a single edge. Next, consider network $G_3$ shown in the Figure \ref{fig:best_c}. As we will show in Proposition~\ref{starmerged} in Section~\ref{sec:charac}, $G_3$ is a best response network to $G_1$.
We have
\begin{align*}
\lim_{n \to \infty} \frac{u(G_3|G_1)}{u(G_2|G_1)}&=\lim_{n \to \infty} \frac{(n-1)(b(1)-c) + (n-2) b(2)}{2(n-2)(b(1)-c)+b(2)}=\infty.
\end{align*}

\begin{figure}[h!]
\begin{minipage}[b]{0.33\textwidth}
\begin{center}
\begin{tikzpicture}
    [scale=.18, inner sep=1pt, minimum size=10pt, auto=center,every node/.style={circle, draw=black, thick}]
  \node (n1) at (11,1)  {\footnotesize $v_1$};
  \node (n2) at (1,1)  {\footnotesize $v_2$};
  \node (n3) at (6,3)  {\footnotesize $v_3$};
  \node (n4) at (6,6)  {\footnotesize $v_4$};

  \node (n6) at (6,11)  {\footnotesize $v_{n}$};
  %\node (n7) at (6,15) {};

  \node [scale=.25, inner sep=1pt, minimum size=1pt, auto=center,every node/.style={circle, draw=black, thick}] (n10) at (6,8)  {};
  \node [scale=.25, inner sep=1pt, minimum size=1pt, auto=center,every node/.style={circle, draw=black, thick}] (n11) at (6,8.5)  {};
  \node [scale=.25, inner sep=1pt, minimum size=1pt, auto=center,every node/.style={circle, draw=black, thick}] (n12) at (6,9)  {};
  \foreach \from/\to in {n1/n2,n1/n3,n2/n3,n1/n4,n2/n4,n1/n6,n2/n6}
    \draw (\from) -- (\to);
\end{tikzpicture}
\end{center}
    \subcaption{$G_1$}
     \label{fig:best_a}
\end{minipage}%
\begin{minipage}[b]{0.33\textwidth}
\begin{center}
\begin{tikzpicture}
    [scale=.18, inner sep=1pt, minimum size=10pt, auto=center,every node/.style={circle, draw=black, thick}]
  \node (n1) at (11,1)  {\footnotesize $v_1$};
  \node (n2) at (1,1)  {\footnotesize $v_2$};
  \node (n3) at (6,3)  {\footnotesize $v_3$};
  \node (n4) at (6,6)  {\footnotesize $v_4$};

  \node (n6) at (6,11)  {\footnotesize $v_{n}$};
  %\node (n7) at (6,15) {};

  \node [scale=.25, inner sep=1pt, minimum size=1pt, auto=center,every node/.style={circle, draw=black, thick}] (n10) at (6,8)  {};
  \node [scale=.25, inner sep=1pt, minimum size=1pt, auto=center,every node/.style={circle, draw=black, thick}] (n11) at (6,8.5)  {};
  \node [scale=.25, inner sep=1pt, minimum size=1pt, auto=center,every node/.style={circle, draw=black, thick}] (n12) at (6,9)  {};
  \foreach \from/\to in {n1/n3,n2/n3,n1/n4,n2/n4,n1/n6,n2/n6}
    \draw (\from) -- (\to);
\end{tikzpicture}
\end{center}
    \subcaption{$G_2$}
 \label{fig:best_b}
\end{minipage}
\begin{minipage}[b]{0.33\textwidth}
\begin{center}
\begin{tikzpicture}
    [scale=.18, inner sep=1pt, minimum size=10pt, auto=center,every node/.style={circle, draw=black, thick}]
  \node (n1) at (11,1)  {\footnotesize $v_1$};
  \node (n2) at (1,1)  {\footnotesize $v_2$};
  \node (n3) at (6,3)  {\footnotesize $v_3$};
  \node (n4) at (6,6)  {\footnotesize $v_4$};

  \node (n6) at (6,11)  {\footnotesize $v_{n}$};

  \node [scale=.25, inner sep=1pt, minimum size=1pt, auto=center,every node/.style={circle, draw=black, thick}] (n10) at (6,8)  {};
  \node [scale=.25, inner sep=1pt, minimum size=1pt, auto=center,every node/.style={circle, draw=black, thick}] (n11) at (6,8.5)  {};
  \node [scale=.25, inner sep=1pt, minimum size=1pt, auto=center,every node/.style={circle, draw=black, thick}] (n12) at (6,9)  {};
  
  \foreach \from/\to in {n1/n2,n2/n3,n2/n4,n2/n6}
    \draw (\from) -- (\to);
\end{tikzpicture}
\end{center}
 \subcaption{$G_3$}
     \label{fig:best_c}
\end{minipage}

\caption{Performance of a greedy algorithm. Graph $G_1$ in (a) is the reference network. Graph $G_2$ in (b) is the output of the greedy algorithm discussed above. Graph $G_3$ in (c) is a best response to $G_1$.}
\label{fig3}
\end{figure}
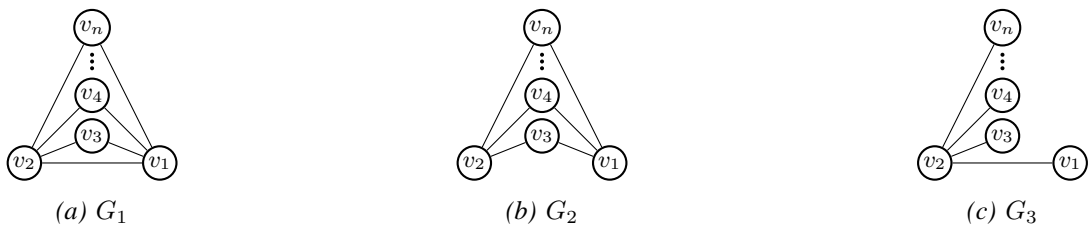

Note that same conclusion is reached even if we start with the complete graph, i.e., we can remove the edges in such a way that we end up in network $G_2$. Thus this greedy algorithm can perform arbitrarily poorly in comparison to the optimal solution.
\end{example}

An important avenue for further research is to find approximation algorithms (and achievable approximation ratios) for the BRN problem.

%%%%%%%%%%%%%%%%%%%%%%%%%%%%%%%%%%%%%%%%%%%%%%%%%%%%%%%%%%%%%%%%%%%%

\subsection{Comparison to Other Network Design Problems}
The problem of optimally designing networks is classical in the computer science and algorithms literature.  Perhaps the most common instance is the minimum spanning tree (MST) problem which is to find a spanning tree of a weighted graph that has the least overall weight; there are greedy algorithms that solve MST in polynomial time \cite{CLRS}.  
Here, we compare the BRN problem to two canonical network design problems that also attempt to minimize distances between pairs of nodes:  the Optimal Communication Spanning Tree (OCST) problem introduced in \cite{Hu}, and the Simple Network Design problem (SNDP) introduced in \cite{johnson}.  

In the OCST problem, for each pair of nodes $v_i,v_j \in N$, there is a {\it communication requirement} $r_{ij} \in \mathbb{N}$.  The goal of the network designer is to construct a tree $T$ on the node set $N$ such that $\sum_{i \ne j}r_{ij}d_T(i,j)$ is minimized.  This problem is polynomial-time solvable for any set of $r_{ij}$ \cite{Hu}.

In the SNDP problem, one is given an undirected graph $G=(N, E)$ and a criterion $C \in \mathbb{N}$.  The objective is to determine if there exists a subgraph $G'=(N, E')$ of $G$ with at most $|N|-1$ edges such that $\sum_{i \ne j}d_{G'}(i,j) \le C$. It was shown in \cite{johnson} that this problem is NP-complete. 

The relationships between the BRN, OCST and SNDP problems are as follows.  
\begin{itemize}
\item The OCST and SNDP problems explicitly constrain the number of edges in the designed network, whereas the BRN problem includes the cost of edges in the utility function.
\item The SNDP problem requires the designed network to be a subgraph of another given network, whereas the BRN and OCST problems place no such constraint.
\item The objective of the SNDP problem is to minimize the sum of distances between {\it all} pairs of nodes, whereas the BRN and OCST problems allow the objective function to only depend on distances between selected pairs of nodes (the OCST problem does this by setting $r_{ij} = 0$ for those pairs that do not wish to communicate).
\end{itemize}
Despite the apparent similarities between the BRN problem and the OCST problem, Theorem~\ref{complexityBR} shows that the BRN problem is NP-hard, even though the OCST problem can be solved in polynomial-time.  This increase in complexity is a byproduct of the additional flexibility afforded by the general nonincreasing benefit function in the BRN problem (as opposed to the scaled distances in the utility function for the OCST problem), which allows it to capture the tree-$t$-spanner problem as a special case.

In the next section of this paper, we will characterize further properties of BR networks; these will allow us to find BR networks with respect to certain specific classes of graphs, which in turn will allow us to formulate and study a multi-layer network formation setting with multiple network designers.

\section{Further Properties of Best Response Networks}
\label{sec:properties}

The proofs of all results in this section are given in Appendix~\ref{app:proof_sectionV}.

\begin{lemma} \label{lem2}
Let $G_2$ be a BR network to $G_1$, and suppose that $G_2$ is not connected. Let $G_{2i}=(N_{i},~E_{2i})$, $i=1, \dots, k$, be the components of $G_2$.  Let $G_{1i}=(N_{i},~E_{1i})$, $i=1,2, \dots, k$, be the subgraphs induced by vertex sets $N_{i}$ on $G_1$. Then network $G_{2i}$ must be a BR network to $G_{1i}$ for $i=1, 2, \dots, k$.
\end{lemma}

The following lemma considers the case when there are isolated nodes in $G_1$.

\begin{lemma}\label{lem4}
Let $G_1 = (N, E_1)$ and suppose $v \in N$ is an isolated node.  Then $v$ is isolated in any BR to $G_1$.
\end{lemma}

The properties described above are independent of the relative values of the benefit function and edge costs.   The following set of results provide more details of the BR networks for certain ranges of benefits and costs.

\begin{lemma}\label{merged}
Let $G_1=(N, E_1)$ be an arbitrary graph. 
\begin{enumerate}
\item If $b(1) - c > b(2)$, then the unique BR network to $G_1$ is $G_2=G_1$.
\item If $b(1)<c$, then $G_1$ is not a BR network to $G_1$, unless $G_1$ is the empty network.
\item Define
\begin{equation}\label{alpha}
\alpha\triangleq \operatorname*{max}_{2 \leq |S|, S \subseteq N}  \frac{|E_{G_1}(S, S)|}{|S|-1}-1,
\end{equation}
where $E_{G_1}(S, S)$ denotes the set of edges in $G_1$ that have both of their endpoints in the set $S$, i.e., $E_{G_1}(S, S)= E_1 \cap (S \times S)$. If $c>b(1)+\alpha b(2)$, then the unique BR network with respect to $G_1$ is the empty network.
\end{enumerate}
\end{lemma}

The parameter $\alpha$ is a measure of the edge density of the underlying graph,\footnote{There exist efficient algorithms to find maximally dense subgraphs in networks \cite{goldberg}.} and thus the threshold to have the empty network as the best response network increases as the underlying graph becomes more dense. The following example illustrates the implication of $\alpha$ for various graphs.

\begin{example} \label{ex:alpha}
In the following, we define $|N|=n$.
\begin{itemize}
\item Assume that $G_1=(N, E_1)$ is the complete graph. Then $|E_{G_1}(S,S)|={|S| \choose 2}$ for any (non-singleton) $S \subseteq N$ and thus $\alpha=\frac{n-2}{2}$ in equation \eqref{alpha}. This means that the BR to the complete graph is the empty graph for $c>b(1)+\frac{n-2}{2} b(2)$, yielding part (iii) of Proposition~\ref{prop:jacksonresult} (obtained in \cite{P4}) as a special case of Lemma~\ref{merged}.

\item Suppose that $G_1=(N, E_1)$ is a tree. Since any induced subgraph of a tree is a forest (it is a tree when the subgraph is connected), we have $|E_{G_1}(S,S)|\leq |S|-1$ for any non-singleton $S \subseteq N$. Thus 
$$
\frac{|E_{G_1}(S, S)|}{|S|-1} -1 \leq 0 ~~~\forall S \subseteq N, ~|S| \geq 2.
$$ 
This means that $\alpha=0$ (which happens for any $S$ that induces a connected subgraph on $G_1$). Therefore, we can conclude that the BR network to a tree is the empty network when $c>b(1)$.

\item Consider a cycle graph $G_1=(N, E_1)$ with $n$ nodes.\footnote{A cycle graph with $n$ nodes consists of only one cycle of length $n$.} Any induced subgraph of $G_1$ on a non-singleton node set $S \subset N$ is an acyclic graph and thus $|E_{G_1}(S,S)| \leq |S|-1$. For $S=N$, we have $|E_{G_1}(N,N)|=n$. Thus $\alpha=\frac{1}{n-1}$, and the BR network to $G_1$ is the empty network for $c>b(1)+\frac{1}{n-1} b(2)$.
\end{itemize}
\end{example}

\begin{remark}
Note that when $b(2) = 0$, Lemma~\ref{merged} indicates that for $b(1) > c$, $G_1$ is a unique BR to itself (for any network $G_1$), and when $b(1) < c$, the empty network is a unique best response (both $G_1$ and the empty network are best responses with utility $0$ when $b(1) = c$).  Thus, in the rest of the paper, we will assume that $b(2) > 0$.
\end{remark}

In the next lemma, we consider the case that we have nodes with degree one in the graph.

\begin{lemma}\label{leafnode}
Let $G_1=(N, E_1)$, and suppose $v \in N$ is a leaf node.  Define the induced subgraph of $G_1$ under the node set $N \setminus \{v\}$ as $G_{11}=(N\setminus\{v\}, E_{11})$ (i.e., the graph obtained by removing node $v$ and its incident edge). Then a BR to $G_1$ can be obtained by first finding a BR to $G_{11}$ and then adding $v$ as an isolated node if $b(1) \leq c$, or adding $v$ together with a single edge to its neighbor in $G_1$ if $b(1) \geq c$. 
\end{lemma}

The above lemma provides the following method to simplify the task of finding a best response network.  Given a graph $G_1$, we recursively remove nodes of degree $1$ until we are left with a graph where all nodes have degree two or larger (this is known as {\it peeling} the graph, and the resulting subgraph is known as a $2$-core \cite{mitzenmacher12}).  A best response to the $2$-core can then be found using whatever means necessary, and then the removed nodes can be recursively added back as isolated nodes (if $b(1) \le c$), or with the single edge that was removed (if $b(1) \geq c$).  

\section{Best Responses to Specific Networks}
\label{sec:charac}
We will now apply the above results to characterize best responses to acyclic networks and networks with a star subgraph.  The latter models, for example, sensor or communication networks where one or more base stations or fusion centers wish to communicate with all nodes, while the other nodes only need to communicate locally amongst themselves. The proofs of the following two propositions are provided in Appendix~\ref{app:proof_forest}.

\begin{proposition} \label{G1forest}
Let $G_1=(N, E_1)$ be a forest. 
\begin{itemize}
\item If $b(1) < c$, the empty network is the unique BR to $G_1$. 
\item If $b(1) > c$, then $G_2=G_1$ is a BR network to $G_1$. 
\item If $b(1)=c$, the empty network and $G_2 = G_1$ are both BR networks to $G_1$. 
\item For $b(1) >\operatorname*{max} \{b(2), c\}$,  the unique BR to $G_1$ is $G_2=G_1$.
\end{itemize}
\end{proposition}

\begin{proposition}\label{starmerged}
Let $G_1=(N, E_1)$ be a graph that has a star subgraph centered at node $v \in N$. 
\begin{itemize}
\item If $b(1)-b(2)>c$, then $G_1$ is the unique BR to $G_1$. 
\item If $b(1)-b(2) \leq c \leq b(1)$, then the star network centered at node $v$ is a BR network to $G_1$.
\item If $b(1) \leq c$, one of the following networks is a BR to $G_1$:
\begin{enumerate}
\item A star network on $N$ with center at node $v$.
\item A network where one component is a star and all other components are isolated nodes.
\item The empty network.
\end{enumerate}
\end{itemize}
\end{proposition}

\section{Multi-Layer Network Formation Game with Strategic Substitutes}
\label{sec:MLNFG} 
In the previous sections, we considered the scenario where a network designer chooses an optimal graph (or layer) $G_2$ with respect to a given graph $G_1$.  In this section, we will build on this formulation to consider a scenario where multiple network designers are building layers, with a utility for each layer that depends on the structure of that layer and the layers constructed by the other designers.  This models, for instance, different mail and courier service companies designing their individual networks to service their customers, or different transportation networks (air, rail, bus) arising between a set of cities \cite{jeanpaul, USPS, porter}.  
We start by defining an $m$-player game where each player corresponds to one of the layers.

\begin{definition}\label{multigame}
A {\it Multi-Layer Network Formation Game} has a set of $m$ players $P = \{P_1, P_2, \dots, P_m\}$. The strategy space for each of the players is defined to be $G^N$ where $N=\{v_1,v_2, \dots, v_n\}$, i.e., the set of all graphs on node set $N$. 
For each $i \in \{1, 2, \ldots, m\}$, let $G_i=(N, E_i) \in G^N$ denote the action of player $P_i$. The utility of player $P_i$ is given by a function $A_i: G^N \times G^N \times \dots \times G^N \to \mathbb{R}$, where the $j^{th}$ argument is the action of the $j^{th}$ player for $1 \leq j \leq m$.
\end{definition}

We will use $G_{-i}$ to denote the vector of actions of all players except player $P_i$, and use $A_i(G_i,G_{-i})$ to denote the utility of player $P_i$ with respect to the given vector $(G_1, G_2, \ldots, G_m)$.  Based on the definition of the game, we say that a vector of networks $(G_1, G_2, \ldots, G_m)$ is a {\it Nash equilibrium} if and only if $G_i\in\operatorname*{arg\,max}_G A_i(G, G_{-i})$ for all $i \in \{1, 2, \ldots, m\}$.  In this case, $G_i$ is said to be a BR network to $G_{-i}$ with respect to the utility function $A_i$.

The characteristics of the game and the optimal strategies for each player will depend on the form of the utility functions $A_i$.  Here, as a starting point for studying such games, we will focus on distance-based utilities (thereby building on our results from the first part of the paper).  The reference networks for the distance-based utility function for each player will depend on the networks constructed by the other players.  
In the remainder of the paper, we will explore functions that view different layers of the network as {\it strategic substitutes}, where the presence of a link in one layer makes it less desirable for that link to appear in another layer; this captures the notion that the different network layers are attempting to fill gaps in connectivity left by the other layers.\footnote{One can also consider a {\it strategic complements} version of this class of games where each player wishes to provide short paths between those pairs of nodes that share an edge in each of the other layers. The analysis of such games is relatively straightforward and thus we focus on strategic substitutes in this paper.}   As a motivating example, consider competing transportation companies offering services between a common set of cities.  
Suppose that for economical reasons, each company would prefer to design its transportation network to provide short routes between those cities that are not directly serviced by any other company.   In other words, each company designs its network with respect to the {\it complement} of the transportation networks provided by all other companies.    
If we impose further structure on such games by assuming distance-based utility functions, we obtain the game defined below.  
In the following definition, for a set of graphs $G_j = (N, E_j)$, $j = 1, 2, \ldots, m$, on a common set of nodes, we use the notation $\cup_{j =1}^mG_j$ to indicate the graph $G = (N, \cup_{j=1}^mE_j)$, and $\cap_{j =1}^mG_j$ to indicate the graph $G = (N, \cap_{j=1}^mE_j)$.

\begin{definition}
The game in Definition \ref{multigame} is said to be a {\it Multi-Layer Network Formation Game with Strategic Substitutes and Distance-Utilities} if the utility functions are of the form
\begin{align} \label{mlnfg}
A_i (G_1, \dots, G_m) &=u_i (G_i|\sim{(\cup_{j=1, j \neq i}^m G_j)})\\ \nonumber
&=\sum_{(x,y) \notin \cup_{k=1, k \neq i}^m {E}_k} b_i(d_{G_i}(x,y))- c_i |E_i|,
\end{align}
where the function $u_i$ is defined in \eqref{P_2}; the benefit functions $b_i(\cdot)$ are nonnegative, nonincreasing and satisfy $b_i(\infty) = 0$, and all costs $c_i$ are positive.  The benefit functions and costs can be different for the different players. 
\label{def:MLNFG_SS_DU}
\end{definition}

It is clear from the definition of the game that $(G_1, G_2, \dots, G_m)$ is a Nash equilibrium if and only if for all $1 \leq i \leq m$, $G_i$ is a BR network with respect to $\sim(\cup_{j=1, j \neq i}^m G_j)$ for the utility function \eqref{P_2}.  
Although we showed in Theorem~\ref{complexityBR} that finding a BR network with respect to this utility function is NP-hard in general, we now show that certain insights can nevertheless be obtained in the multiplayer setting (regardless of the number of nodes and players).   To develop our results, we partition the set of players $P$ into three sets: {\it high-cost players} $S_H = \{P_i \in P | c_i > b_i(1)\}$, {\it medium-cost players} $S_M = \{P_i \in P | b_i(1) \ge c_i \ge b_i(1)-b_i(2)\}$ and {\it low-cost players} $S_L = \{P_i \in P |b_i(1)-b_i(2)>c_i\}$.     
We start by considering the case where the game contains low-cost players.

%%%%%%%%%%%%%%%%%%%%%%%%%%%%%%%%%%%%%%%%%%%%%%%
%%%%%%%%%%%%%%%%%%%%%%%%%%%%%%%%%%%%%%%%%%%%%%%%
\subsection{Games Containing Low-Cost Players}

\begin{proposition}
Suppose $|S_L| \ge 1$.  Then in every Nash equilibrium, every player in $S_H$ chooses the empty network.  Furthermore, any vector of disjoint networks $(G_1, G_2, \ldots, G_m)$ forms a Nash equilibrium when $\{G_k | P_k \in S_M\}$ is a set of disjoint forests and $\cup_{i \in S_L}G_i = {\sim\cup_{i \in S_M}G_i}$.
\label{prop:low_cost_player}
\end{proposition}

\begin{IEEEproof}
Let $(G_1, G_2, \ldots, G_m)$ be any vector of networks in Nash equilibrium.  Since there exists at least one player $P_i$ whose edge cost satisfies $c_i < b_i(1)-b_i(2)$, the Nash equilibrium vector must satisfy $\cup_{j = 1}^m G_j = G^c$, where $G^c$ is the complete network.  To see this, suppose that the union of the graphs is not the complete network; then there exists some edge $(u,v)$ that does not appear in any network, and thus appears in the complement of the graph $\cup_{j = 1, j \ne i}^m G_j$.  By Lemma~\ref{merged}, the BR to $\sim\cup_{j = 1, j \ne i}^m G_j$ with respect to player $P_i$'s utility function is $\sim\cup_{j = 1, j \ne i}^m G_j$, and thus the edge $(u,v)$ appears in graph $G_i$, contradicting the fact that it does not appear in the union of all the graphs.

Next, note that since $\cup_{j = 1}^m G_j = G^c$, for any player $P_k \in P$, the graph $G_k = (N, E_k)$ is a BR to the graph $G^c \setminus \{\cup_{j = 1, j \ne k}^m G_j\} \subseteq G_k$.  By Lemma~\ref{lem1}, a BR to a graph cannot be a strict superset of that graph, and thus we have that $G_k$ is a best response to itself with respect to the utility function of player $P_k$.  Now if $P_k \in S_H$, we know from Lemma~\ref{merged} that $G_k$ must be the empty network, completing the first part of the proof.  For the second part, note that for any vector of networks satisfying the given properties,  Proposition~\ref{G1forest} and Lemma~\ref{merged} indicate that a best response to $G_k$ is indeed $G_k$ for $P_k \in S_M \cup S_L$, completing the proof.
\end{IEEEproof}

The above result shows that the presence of a player with low edge costs (relative to its own benefit function) guarantees the existence of a Nash equilibrium in the game, and furthermore, such low-cost players drive players with sufficiently high edge costs out of the game; the proposition provides the threshold for costs at which this occurs (namely $b_i(1) < c_i$).  Players with medium edge costs, on the other hand, can obtain certain nonempty networks in equilibrium, and the players with low edge costs split all of the remaining edges amongst themselves.  

We now study the situation where there are no low-cost players in the game (i.e., $S_L = \emptyset$).  We start by considering games that contain only high-cost players.

%%%%%%%%%%%%%%%%%%%%%%%%%%%%%%%%%%%%%%%%%%%%%%%
%%%%%%%%%%%%%%%%%%%%%%%%%%%%%%%%%%%%%%%%%%%%%%%%
\subsection{Games Containing Only High-Cost Players}

Suppose $P = S_H$ (i.e., $S_L = S_M = \emptyset$).  For each player $P_i \in P$, define the index $k_i$ as
$$
k_i \triangleq \min \left\{ t \in \mathbb{N} \mid c_i < b_i(1)+\frac{t-2}{2} b_i(2)\right\}.
$$
Since $c_i >b_i(1)$, we have $k_i \geq 3$ for all $P_i \in P$. If $k_i > n$, by Lemma~\ref{merged}, the empty network is a BR of player $P_i$ to any set of networks $G_{-i}$ (since $\alpha$ in \eqref{alpha} satisfies $\alpha \le \frac{n-2}{2}$ for any reference graph). Thus without loss of generality, assume that all players have $3 \leq k_i \leq n$ and players are sorted according to their $k_i$, i.e., $k_1 \leq k_2 \leq \cdots \leq k_m \leq n$.  
We will now partition the set of players $P$ into different sets.

Define the index $i_1$ as
$$
i_1 \triangleq \max \left\{ i \in \{1, 2, \ldots, m\} \mid k_{i} \le n-i+1 \right\}.
$$
Next, define
\begin{align}\nonumber
i_2 &\triangleq \max \left\{ i \in \{1, 2, \ldots, i_1 -1\} \mid k_{i} \le i_1-i+1 \right\},\\ \nonumber
i_3 &\triangleq \max \left\{ i \in \{1, 2, \ldots, i_2 -1 \} \mid k_{i} \le i_2-i+1 \right\},\\ \label{i_r}
&\enspace \vdots \\  
i_r &\triangleq \max \left\{ i \in \{1, 2, \ldots, i_{r-1} - 1\} \mid k_{i} \le i_{r-1}-i+1 \right\}, \nonumber
\end{align}
where $i_r$ satisfies $i_r < k_1$ (so that no further sets of this form can be defined).

The above indices satisfy $1 \le i_r < i_{r-1} < \cdots < i_1 \le m$.  Partition the set of players and nodes as follows
\begin{align} \nonumber
H_r &= \{P_1,\ldots,P_{ i_r}\}, V_r =\{v_1,\ldots, v_{i_r}\}\\  \nonumber
H_{r-1} &=\{P_{i_{r}+1},\ldots,P_{ i_{r-1}}\}, V_{r-1}=\{v_{i_{r}+1},\ldots,v_{i_{r-1}}\}\\ \label{eqn:PV}
&\enspace \vdots  \\ \nonumber
H_1 &= \{P_{i_2+1},\ldots,P_{i_1}\}, V_1=\{v_{i_2+1},\ldots,v_{i_1}\}. 
\end{align}
Also define $H_0 = \{P_{i_1 + 1}, P_{i_1 + 2}, \ldots, P_m\}$ and $V_0=\{v_{i_1+1},v_{i_1 + 2}, \ldots,v_n\}$.   
\begin{proposition}
For each player $P_j \in H_l$ (for $1 \le l \le r$), define the network $G_j$ to be the star network centered on node $v_j$ with peripheral nodes $\cup_{t=0}^{l-1} V_t$, where $H_l$ and $V_t$ are defined as in equation \eqref{eqn:PV}.  For each player $P_j \in H_0$, define $G_j$ to be the empty network.  Then the set of networks $(G_1, G_2, \ldots, G_m)$ forms a Nash equilibrium.
\label{prop:high_cost_player}
\end{proposition}

The proof of the above proposition is given in Appendix~\ref{app:proof_high_cost}.  The following example illustrates the structure of the Nash equilibrium specified by the above proposition.

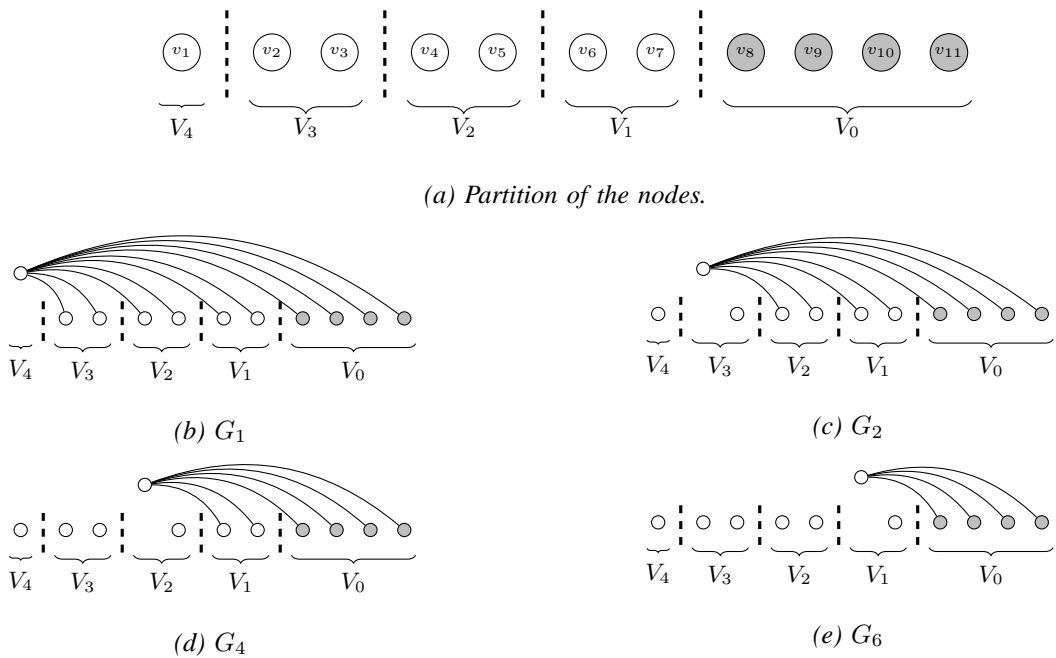
\begin{figure}[ht!]
\begin{minipage}{0.98\textwidth}
\begin{center}
\begin{tikzpicture}
  %[scale=.24, inner sep=1pt, minimum size=0.6cm, auto=center, node/.style={circle, draw=black, thick}]
  [scale=.3, inner sep=1pt, minimum size=14pt, auto=center, node/.style={circle, draw=black, thick}]
  \node [circle, draw, fill=white](n1) at (-3,0)  {\tiny $v_1$};

  \node [circle, draw, fill=white](n2) at (1,0)  {\tiny $v_2$};
  \node [circle, draw, fill=white](n3) at (4,0)  {\tiny $v_3$};

  \node [circle, draw, fill=white](n4) at (8,0)  {\tiny $v_4$};
  \node [circle, draw, fill=white](n5) at (11,0)  {\tiny $v_5$};

  \node [circle, draw, fill=white](n6) at (15,0) {\tiny $v_6$};
  \node [circle, draw, fill=white](n7) at (18,0) {\tiny $v_7$};

  \node [circle, draw, fill=gray!50](n8) at (22,0) {\tiny $v_8$};
  \node [circle, draw, fill=gray!50](n9) at (25,0) {\tiny $v_9$};
  \node [circle, draw, fill=gray!50](n10) at (28,0) {\tiny $v_{10}$};
  \node [circle, draw, fill=gray!50](n11) at (31,0) {\tiny $v_{11}$};

%  \foreach \from/\to in {n1/n3,n1/n4,n1/n5,n1/n6,n1/n7,n8/n1, n1/n9, n1/n10, n1/n11}
   % \draw[bend] (\from) -- (\to);
  \draw[very thick, black, dashed] 
    (-1,-2)--(-1,2)
    (6,-2)--(6,2)
    (13,-2)--(13,2)
    (20,-2)--(20,2);

\draw [decorate,decoration={brace,amplitude=1.5pt,mirror,raise=0.5pt},yshift=0pt]
(-4,-2.2) -- (-2,-2.2) node [black,midway,yshift=-0.35cm] {\footnotesize $V_4$};

\draw [decorate,decoration={brace,amplitude=5pt,mirror,raise=1pt},yshift=0pt]
(0,-2) -- (5,-2) node [black,midway,yshift=-0.4cm] {\footnotesize $V_3$};

\draw [decorate,decoration={brace,amplitude=5pt,mirror,raise=1pt},yshift=0pt]
(7,-2) -- (12,-2) node [black,midway,yshift=-0.4cm] {\footnotesize $V_2$};

\draw [decorate,decoration={brace,amplitude=5pt,mirror,raise=1pt},yshift=0pt]
(14,-2) -- (19,-2) node [black,midway,yshift=-0.4cm] {\footnotesize $V_1$};

\draw [decorate,decoration={brace,amplitude=5pt,mirror,raise=1pt},yshift=0pt]
(21,-2) -- (32,-2) node [black,midway,yshift=-0.4cm] {\footnotesize $V_0$};

%\node[legend_isps] (i1) at (-1,-4) {\textsc{$V_1$}};

\end{tikzpicture}
\end{center}
\subcaption{Partition of the nodes. }
\label{fig_struct_1}
\end{minipage}

\begin{minipage}{0.46\textwidth}
\begin{center}
\begin{tikzpicture}
  [scale=.15, inner sep=1pt, minimum size=5pt, auto=center, node/.style={circle, draw=black, thick}]
  \node [circle, draw, fill=white](n1) at (-3,4)  {};

  \node [circle, draw, fill=white](n2) at (1,0)  {};
  \node [circle, draw, fill=white](n3) at (4,0)  {};

  \node [circle, draw, fill=white](n4) at (8,0)  {};
  \node [circle, draw, fill=white](n5) at (11,0)  {};

  \node [circle, draw, fill=white](n6) at (15,0) {};
  \node [circle, draw, fill=white](n7) at (18,0) {};

  \node [circle, draw, fill=gray!50](n8) at (22,0) {};
  \node [circle, draw, fill=gray!50](n9) at (25,0) {};
  \node [circle, draw, fill=gray!50](n10) at (28,0) {};
  \node [circle, draw, fill=gray!50](n11) at (31,0) {};

%  \foreach \from/\to in {n1/n3,n1/n4,n1/n5,n1/n6,n1/n7,n8/n1, n1/n9, n1/n10, n1/n11}
   % \draw[bend] (\from) -- (\to);
  \draw[very thick, black, dashed] 
    (-1,-2)--(-1,2)
    (6,-2)--(6,2)
    (13,-2)--(13,2)
    (20,-2)--(20,2);

\draw [decorate,decoration={brace,amplitude=1.5pt,mirror,raise=0.5pt},yshift=0pt]
(-4,-2.2) -- (-2,-2.2) node [black,midway,yshift=-0.35cm] {\footnotesize $V_4$};

\draw [decorate,decoration={brace,amplitude=3pt,mirror,raise=1pt},yshift=0pt]
(0,-2) -- (5,-2) node [black,midway,yshift=-0.4cm] {\footnotesize $V_3$};

\draw [decorate,decoration={brace,amplitude=3pt,mirror,raise=1pt},yshift=0pt]
(7,-2) -- (12,-2) node [black,midway,yshift=-0.4cm] {\footnotesize $V_2$};

\draw [decorate,decoration={brace,amplitude=3pt,mirror,raise=1pt},yshift=0pt]
(14,-2) -- (19,-2) node [black,midway,yshift=-0.4cm] {\footnotesize $V_1$};

\draw [decorate,decoration={brace,amplitude=3pt,mirror,raise=1pt},yshift=0pt]
(21,-2) -- (32,-2) node [black,midway,yshift=-0.4cm] {\footnotesize $V_0$};

\path[every node/.style={font=\sffamily\small}]
(n1) edge [bend left] node[left] {} (n2)
	 edge [bend left] node[left] {} (n3)
	 edge [bend left] node[left] {} (n4)
	 edge [bend left] node[left] {} (n5)
	 edge [bend left] node[left] {} (n6)
	 edge [bend left] node[left] {} (n7)
	 edge [bend left] node[left] {} (n8)
	 edge [bend left] node[left] {} (n9)
	 edge [bend left] node[left] {} (n10)
	 edge [bend left] node[left] {} (n11);

\end{tikzpicture}
\end{center}
\subcaption{$G_1$}
\label{fig_struct_2}
\end{minipage}
\begin{minipage}{0.46\textwidth}
\begin{center}
\begin{tikzpicture}
  [scale=.15, inner sep=1pt, minimum size=5pt, auto=center, node/.style={circle, draw=black, thick}]
  \node [circle, draw, fill=white](n1) at (-3,0)  {};

  \node [circle, draw, fill=white](n2) at (1,4)  {};
  \node [circle, draw, fill=white](n3) at (4,0)  {};

  \node [circle, draw, fill=white](n4) at (8,0)  {};
  \node [circle, draw, fill=white](n5) at (11,0)  {};

  \node [circle, draw, fill=white](n6) at (15,0) {};
  \node [circle, draw, fill=white](n7) at (18,0) {};

  \node [circle, draw, fill=gray!50](n8) at (22,0) {};
  \node [circle, draw, fill=gray!50](n9) at (25,0) {};
  \node [circle, draw, fill=gray!50](n10) at (28,0) {};
  \node [circle, draw, fill=gray!50](n11) at (31,0) {};

%  \foreach \from/\to in {n1/n3,n1/n4,n1/n5,n1/n6,n1/n7,n8/n1, n1/n9, n1/n10, n1/n11}
   % \draw[bend] (\from) -- (\to);
  \draw[very thick, black, dashed] 
    (-1,-2)--(-1,2)
    (6,-2)--(6,2)
    (13,-2)--(13,2)
    (20,-2)--(20,2);

\draw [decorate,decoration={brace,amplitude=1.5pt,mirror,raise=0.5pt},yshift=0pt]
(-4,-2.2) -- (-2,-2.2) node [black,midway,yshift=-0.35cm] {\footnotesize $V_4$};

\draw [decorate,decoration={brace,amplitude=3pt,mirror,raise=1pt},yshift=0pt]
(0,-2) -- (5,-2) node [black,midway,yshift=-0.4cm] {\footnotesize $V_3$};

\draw [decorate,decoration={brace,amplitude=3pt,mirror,raise=1pt},yshift=0pt]
(7,-2) -- (12,-2) node [black,midway,yshift=-0.4cm] {\footnotesize $V_2$};

\draw [decorate,decoration={brace,amplitude=3pt,mirror,raise=1pt},yshift=0pt]
(14,-2) -- (19,-2) node [black,midway,yshift=-0.4cm] {\footnotesize $V_1$};

\draw [decorate,decoration={brace,amplitude=3pt,mirror,raise=1pt},yshift=0pt]
(21,-2) -- (32,-2) node [black,midway,yshift=-0.4cm] {\footnotesize $V_0$};

\path[every node/.style={font=\sffamily\small}]
(n2) edge [bend left] node[left] {} (n4)
	 edge [bend left] node[left] {} (n5)
	 edge [bend left] node[left] {} (n6)
	 edge [bend left] node[left] {} (n7)
	 edge [bend left] node[left] {} (n8)
	 edge [bend left] node[left] {} (n9)
	 edge [bend left] node[left] {} (n10)
	 edge [bend left] node[left] {} (n11);

\end{tikzpicture}
\end{center}
\subcaption{$G_2$}
\label{fig_struct_3}
\end{minipage}

\begin{minipage}{0.46\textwidth}
\begin{center}
\begin{tikzpicture}
  [scale=.15, inner sep=1pt, minimum size=5pt, auto=center, node/.style={circle, draw=black, thick}]
  \node [circle, draw, fill=white](n1) at (-3,0)  {};

  \node [circle, draw, fill=white](n2) at (1,0)  {};
  \node [circle, draw, fill=white](n3) at (4,0)  {};

  \node [circle, draw, fill=white](n4) at (8,4)  {};
  \node [circle, draw, fill=white](n5) at (11,0)  {};

  \node [circle, draw, fill=white](n6) at (15,0) {};
  \node [circle, draw, fill=white](n7) at (18,0) {};

  \node [circle, draw, fill=gray!50](n8) at (22,0) {};
  \node [circle, draw, fill=gray!50](n9) at (25,0) {};
  \node [circle, draw, fill=gray!50](n10) at (28,0) {};
  \node [circle, draw, fill=gray!50](n11) at (31,0) {};

%  \foreach \from/\to in {n1/n3,n1/n4,n1/n5,n1/n6,n1/n7,n8/n1, n1/n9, n1/n10, n1/n11}
   % \draw[bend] (\from) -- (\to);
  \draw[very thick, black, dashed] 
    (-1,-2)--(-1,2)
    (6,-2)--(6,2)
    (13,-2)--(13,2)
    (20,-2)--(20,2);

\draw [decorate,decoration={brace,amplitude=1.5pt,mirror,raise=0.5pt},yshift=0pt]
(-4,-2.2) -- (-2,-2.2) node [black,midway,yshift=-0.35cm] {\footnotesize $V_4$};

\draw [decorate,decoration={brace,amplitude=3pt,mirror,raise=1pt},yshift=0pt]
(0,-2) -- (5,-2) node [black,midway,yshift=-0.4cm] {\footnotesize $V_3$};

\draw [decorate,decoration={brace,amplitude=3pt,mirror,raise=1pt},yshift=0pt]
(7,-2) -- (12,-2) node [black,midway,yshift=-0.4cm] {\footnotesize $V_2$};

\draw [decorate,decoration={brace,amplitude=3pt,mirror,raise=1pt},yshift=0pt]
(14,-2) -- (19,-2) node [black,midway,yshift=-0.4cm] {\footnotesize $V_1$};

\draw [decorate,decoration={brace,amplitude=3pt,mirror,raise=1pt},yshift=0pt]
(21,-2) -- (32,-2) node [black,midway,yshift=-0.4cm] {\footnotesize $V_0$};

\path[every node/.style={font=\sffamily\small}]
(n4) edge [bend left] node[left] {} (n6)
	 edge [bend left] node[left] {} (n7)
	 edge [bend left] node[left] {} (n8)
	 edge [bend left] node[left] {} (n9)
	 edge [bend left] node[left] {} (n10)
	 edge [bend left] node[left] {} (n11);

\end{tikzpicture}
\end{center}
\subcaption{$G_4$}
\label{fig_struct_4}
\end{minipage}
\begin{minipage}{0.46\textwidth}
\begin{center}
\begin{tikzpicture}
   [scale=.15, inner sep=1pt, minimum size=5pt, auto=center, node/.style={circle, draw=black, thick}]
  \node [circle, draw, fill=white](n1) at (-3,0)  {};

  \node [circle, draw, fill=white](n2) at (1,0)  {};
  \node [circle, draw, fill=white](n3) at (4,0)  {};

  \node [circle, draw, fill=white](n4) at (8,0)  {};
  \node [circle, draw, fill=white](n5) at (11,0)  {};

  \node [circle, draw, fill=white](n6) at (15,4) {};
  \node [circle, draw, fill=white](n7) at (18,0) {};

  \node [circle, draw, fill=gray!50](n8) at (22,0) {};
  \node [circle, draw, fill=gray!50](n9) at (25,0) {};
  \node [circle, draw, fill=gray!50](n10) at (28,0) {};
  \node [circle, draw, fill=gray!50](n11) at (31,0) {};

%  \foreach \from/\to in {n1/n3,n1/n4,n1/n5,n1/n6,n1/n7,n8/n1, n1/n9, n1/n10, n1/n11}
   % \draw[bend] (\from) -- (\to);
  \draw[very thick, black, dashed] 
    (-1,-2)--(-1,2)
    (6,-2)--(6,2)
    (13,-2)--(13,2)
    (20,-2)--(20,2);

\draw [decorate,decoration={brace,amplitude=1.5pt,mirror,raise=0.5pt},yshift=0pt]
(-4,-2.2) -- (-2,-2.2) node [black,midway,yshift=-0.35cm] {\footnotesize $V_4$};

\draw [decorate,decoration={brace,amplitude=3pt,mirror,raise=1pt},yshift=0pt]
(0,-2) -- (5,-2) node [black,midway,yshift=-0.4cm] {\footnotesize $V_3$};

\draw [decorate,decoration={brace,amplitude=3pt,mirror,raise=1pt},yshift=0pt]
(7,-2) -- (12,-2) node [black,midway,yshift=-0.4cm] {\footnotesize $V_2$};

\draw [decorate,decoration={brace,amplitude=3pt,mirror,raise=1pt},yshift=0pt]
(14,-2) -- (19,-2) node [black,midway,yshift=-0.4cm] {\footnotesize $V_1$};

\draw [decorate,decoration={brace,amplitude=3pt,mirror,raise=1pt},yshift=0pt]
(21,-2) -- (32,-2) node [black,midway,yshift=-0.4cm] {\footnotesize $V_0$};

\path[every node/.style={font=\sffamily\small}]
(n6) edge [bend left] node[left] {} (n8)
	 edge [bend left] node[left] {} (n9)
	 edge [bend left] node[left] {} (n10)
	 edge [bend left] node[left] {} (n11);

\end{tikzpicture}
\end{center}
\subcaption{$G_6$}
\label{fig_struct_5}
\end{minipage}

\caption{A multi-layer network formation game considered in Example~\ref{ex:high_cost_game} with $9$ high-cost players and $11$ nodes.  Nodes are partitioned into 5 sets as shown in Figure \ref{fig_struct_1}, based on the characteristics of the players.  Each node in each of the sets $V_1, V_2, V_3, V_4$ will be chosen by a different player as the center of a star subgraph in the Nash equilibrium. The Nash equilibrium networks of players $P_1$, $P_2$, $P_4$ and $P_6$ are shown in \ref{fig_struct_2}, \ref{fig_struct_3}, \ref{fig_struct_4} and \ref{fig_struct_5}, respectively.  The networks of players $P_3$, $P_5$ and $P_7$ are not shown; they have stars centered on $v_3$, $v_5$ and $v_7$, respectively, with the same peripheral nodes as $P_2$, $P_4$ and $P_6$, respectively. Players $P_8$ and $P_9$ choose the empty network.}
\label{fig6}
\end{figure}

\begin{example}
Suppose that there are 11 nodes and 9 high-cost players with $k_i=3$ for $1 \leq i \leq 5$, $k_6=4, ~k_7=5$ and $k_8, k_9 \geq 5$. From the equations in \eqref{i_r}, we get $i_1=7, i_2=5, i_3=3$ and $i_4 = 1$. Figure~\ref{fig6} demonstrates the networks of players $P_1$, $P_2$, $P_4$ and $P_6$ in the Nash equilibrium defined in Proposition \ref{prop:high_cost_player}. Player $P_3$ has a similar network to player $P_2$ (except that the star of her network is centered on $v_3$).  Players $P_5$ and $P_7$ have similar networks to that of $P_4$ and $P_6$, respectively (the only difference being that player $P_5$ has a star centered on $v_5$, and $P_7$ has a star centered on $v_7$).  Players $P_8$ and $P_9$ each have the empty network.
\label{ex:high_cost_game}
\end{example}

Despite the stylized nature of the multi-layer network formation game in Definition~\ref{def:MLNFG_SS_DU}, it is of interest to note that the ``hub-and-spoke'' networks that arise in the above Nash equilibrium are predominant in real-world transportation systems (airline networks, in particular) \cite{jeanpaul, USPS, porter}.  While previous work has shown that such networks are optimal in the single-layer setting (e.g., Proposition~\ref{prop:jacksonresult} \cite{P4}), our analysis shows that these structures also arise when players selfishly optimize their individual networks in competitive environments.  
We will now consider games with a mix of medium-cost and high-cost players, and show that such structures also arise as a Nash equilibrium in that setting.

%%%%%%%%%%%%%%%%%%%%%%%%%%%%%%%%%%%%%%%%%%%%%%%
%%%%%%%%%%%%%%%%%%%%%%%%%%%%%%%%%%%%%%%%%%%%%%%%
\subsection{Games With Medium and High-Cost Players}

\begin{proposition}
Suppose that $S_L = \emptyset$, and assume without loss of generality that the first $\mu$ players in $P$ are medium-cost players, with $1\le \mu \le n$.   For $j \in \{1, 2, \ldots, \mu\}$, define the network $G_j$ to be the star network centered on node $v_j$ with peripheral nodes $\{v_{j+1}, v_{j+2}, \ldots, v_n\}$.  For the set of high-cost players $S_H$, let $(G_{\mu+1}, G_{\mu + 2}, \ldots, G_{m})$ be the Nash equilibrium networks on node set $\{v_{\mu+1}, v_{\mu + 2}, \ldots, v_n\}$ defined in Proposition~\ref{prop:high_cost_player}.  Then the set of networks $(G_1, G_2, \ldots, G_m)$ is a Nash equilibrium.
\label{prop:med_cost_player}
\end{proposition}

\begin{IEEEproof}
In the proof, we will use the fact that each network $G_j$, $1 \le j \le m$, only contains edges from node $v_j$ to nodes with index larger than $j$.  For each player $P_j$, let $G_{j,ref} \triangleq \bigcup_{i = 1, i \ne j}^mG_j$ be the union of the networks of the other players.

Consider a medium-cost player $P_j$, where $j \in \{1, 2, \ldots, \mu\}$.    Since all players with index smaller than $j$ are medium-cost players, for each node $v_i$ with $i < j$, $G_{j,ref}$ contains an edge from node $v_i$ to $v_k$ for all $k > i$.  Furthermore, $G_{j,ref}$ contains no edge from $v_k$ to $v_j$ for any $k > j$.   Thus, in the network $\sim G_{j,ref}$, nodes $v_1, v_2, \ldots, v_{j-1}$ are isolated, and there is an edge from $v_j$ to each node $v_k$ with $k > j$.  By Lemma~\ref{lem4}, the isolated nodes in $\sim G_{j, ref}$ remain isolated in the BR; applying Proposition~\ref{starmerged}, a star network centered at $v_j$ with edges to $\{v_{j+1}, \ldots, v_n\}$ is a BR with respect to $\sim G_{j,ref}$.  Thus, $G_j$ is a BR to $\sim G_{j,ref}$.  

Now consider a high-cost player $P_j$, where $j \in \{\mu+1, \mu+2, \ldots, m\}$.  Arguing as above, nodes $v_1, v_2, \ldots, v_\mu$ are isolated in the network $\sim G_{j, ref}$.  Thus by Lemma~\ref{lem4}, those nodes remain isolated in the BR to $\sim G_{j, ref}$.  Since this is true for all high-cost players, we can remove the nodes $v_1, v_2, \ldots, v_{\mu}$ from consideration, and focus on showing that the subgraph of $G_j$ induced by the node set $\{v_{\mu+1}, v_{\mu+2}, \ldots, v_{n}\}$ is a BR to the graphs $(G_{\mu+1}, G_{\mu+2}, \ldots, G_{m})$ on that node set.  This is true by construction, and thus 
the given set of networks is a Nash equilibrium.
\end{IEEEproof}

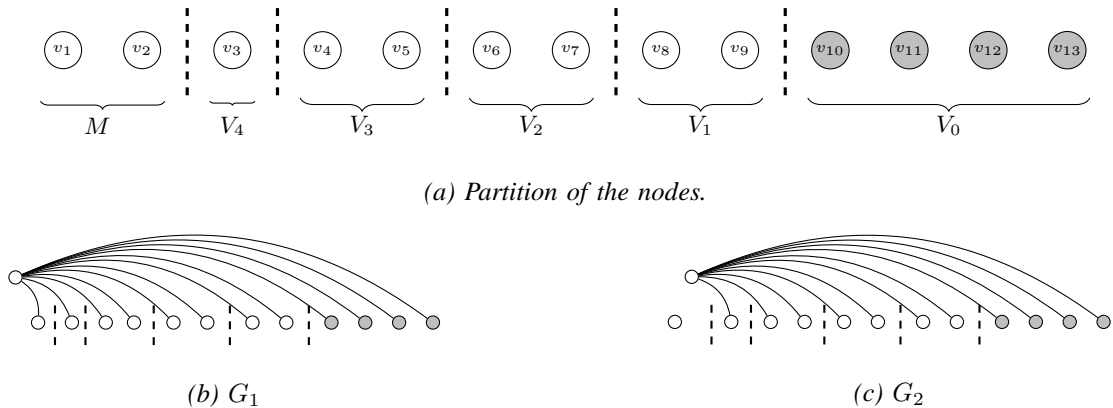
\begin{figure}[ht!]
\begin{minipage}{0.98\textwidth}
\begin{center}
\begin{tikzpicture}
  [scale=.3, inner sep=1pt, minimum size=14pt, auto=center, node/.style={circle, draw=black, thick}]
  \node [circle, draw, fill=white](n1) at (-10.5,0)  {\tiny $v_1$};
  \node [circle, draw, fill=white](n2) at (-7,0)  {\tiny $v_2$};

  \node [circle, draw, fill=white](n3) at (-3,0)  {\tiny $v_3$};

  \node [circle, draw, fill=white](n4) at (1,0)  {\tiny $v_4$};
  \node [circle, draw, fill=white](n5) at (4.5,0)  {\tiny $v_5$};

  \node [circle, draw, fill=white](n6) at (8.5,0)  {\tiny $v_6$};
  \node [circle, draw, fill=white](n7) at (12,0)  {\tiny $v_7$};

  \node [circle, draw, fill=white](n8) at (16,0) {\tiny $v_8$};
  \node [circle, draw, fill=white](n9) at (19.5,0) {\tiny $v_9$};

  \node [circle, draw, fill=gray!50](n10) at (23.5,0) {\tiny $v_{10}$};
  \node [circle, draw, fill=gray!50](n11) at (27,0) {\tiny $v_{11}$};
  \node [circle, draw, fill=gray!50](n12) at (30.5,0) {\tiny $v_{12}$};
  \node [circle, draw, fill=gray!50](n13) at (34,0) {\tiny $v_{13}$};

  \draw[very thick, black, dashed] 
    (-5,-2)--(-5,2)
    (-1,-2)--(-1,2)
    (6.5,-2)--(6.5,2)
    (14,-2)--(14,2)
    (21.5,-2)--(21.5,2);

\draw [decorate,decoration={brace,amplitude=3pt,mirror,raise=0.5pt},yshift=0pt]
(-11.5,-2.2) -- (-6,-2.2) node [black,midway,yshift=-0.35cm, xshift=-2] {\footnotesize $M$};

\draw [decorate,decoration={brace,amplitude=1.5pt,mirror,raise=0.5pt},yshift=0pt]
(-4,-2.2) -- (-2,-2.2) node [black,midway,yshift=-0.35cm] {\footnotesize $V_4$};

\draw [decorate,decoration={brace,amplitude=5pt,mirror,raise=1pt},yshift=0pt]
(0,-2) -- (5.5,-2) node [black,midway,yshift=-0.4cm] {\footnotesize $V_3$};

\draw [decorate,decoration={brace,amplitude=5pt,mirror,raise=1pt},yshift=0pt]
(7.5,-2) -- (13,-2) node [black,midway,yshift=-0.4cm] {\footnotesize $V_2$};

\draw [decorate,decoration={brace,amplitude=5pt,mirror,raise=1pt},yshift=0pt]
(15,-2) -- (20.5,-2) node [black,midway,yshift=-0.4cm] {\footnotesize $V_1$};

\draw [decorate,decoration={brace,amplitude=5pt,mirror,raise=1pt},yshift=0pt]
(22.5,-2) -- (35,-2) node [black,midway,yshift=-0.4cm] {\footnotesize $V_0$};

\end{tikzpicture}
\end{center}
\subcaption{Partition of the nodes. }
\label{fig_med_1}
\end{minipage}

\begin{minipage}{0.48\textwidth}
\begin{center}
\begin{tikzpicture}
  [scale=.15, inner sep=1pt, minimum size=5pt, auto=center, node/.style={circle, draw=black, thick}]
  \node [circle, draw, fill=white](n1) at (-6,4)  {};
  \node [circle, draw, fill=white](n2) at (-4,0)  {};

  \node [circle, draw, fill=white](n3) at (-1,0)  {};

  \node [circle, draw, fill=white](n4) at (2,0)  {};
  \node [circle, draw, fill=white](n5) at (4.5,0)  {};

  \node [circle, draw, fill=white](n6) at (8,0)  {};
  \node [circle, draw, fill=white](n7) at (11,0)  {};

  \node [circle, draw, fill=white](n8) at (15,0) {};
  \node [circle, draw, fill=white](n9) at (18,0) {};

  \node [circle, draw, fill=gray!50](n10) at (22,0) {};
  \node [circle, draw, fill=gray!50](n11) at (25,0) {};
  \node [circle, draw, fill=gray!50](n12) at (28,0) {};
  \node [circle, draw, fill=gray!50](n13) at (31,0) {};

  \draw[thick, black, dashed] 
    (-2.5,-2)--(-2.5,2)
    (0.2,-2)--(0.2,2)
    (6.25,-2)--(6.25,2)
    (13,-2)--(13,2)
    (20,-2)--(20,2);

\path[every node/.style={font=\sffamily\small}]
(n1) edge [bend left] node[left] {} (n2)
	 edge [bend left] node[left] {} (n3)
	 edge [bend left] node[left] {} (n4)
	 edge [bend left] node[left] {} (n5)
	 edge [bend left] node[left] {} (n6)
	 edge [bend left] node[left] {} (n7)
	 edge [bend left] node[left] {} (n8)
	 edge [bend left] node[left] {} (n9)
	 edge [bend left] node[left] {} (n10)
	 edge [bend left] node[left] {} (n11)
	 edge [bend left] node[left] {} (n12)
	 edge [bend left] node[left] {} (n13);

\end{tikzpicture}
\end{center}
\subcaption{$G_1$}
\label{fig_med_2}
\end{minipage}
\begin{minipage}{0.48\textwidth}
\begin{center}
\begin{tikzpicture}
  [scale=.15, inner sep=1pt, minimum size=5pt, auto=center, node/.style={circle, draw=black, thick}]
  \node [circle, draw, fill=white](n1) at (-7,0)  {};
  \node [circle, draw, fill=white](n2) at (-5.5,4)  {};

  \node [circle, draw, fill=white](n3) at (-2,0)  {};

  \node [circle, draw, fill=white](n4) at (1.5,0)  {};
  \node [circle, draw, fill=white](n5) at (4.5,0)  {};

  \node [circle, draw, fill=white](n6) at (8,0)  {};
  \node [circle, draw, fill=white](n7) at (11,0)  {};

  \node [circle, draw, fill=white](n8) at (15,0) {};
  \node [circle, draw, fill=white](n9) at (18,0) {};

  \node [circle, draw, fill=gray!50](n10) at (22,0) {};
  \node [circle, draw, fill=gray!50](n11) at (25,0) {};
  \node [circle, draw, fill=gray!50](n12) at (28,0) {};
  \node [circle, draw, fill=gray!50](n13) at (31,0) {};

 \draw[thick, black, dashed] 
    (-3.75,-2)--(-3.75,2)
    (-0.25,-2)--(-0.25,2)
    (6.25,-2)--(6.25,2)
    (13,-2)--(13,2)
    (20,-2)--(20,2);

\path[every node/.style={font=\sffamily\small}]
(n2) edge [bend left] node[left] {} (n3)
	 edge [bend left] node[left] {} (n4)
	 edge [bend left] node[left] {} (n5)
	 edge [bend left] node[left] {} (n6)
	 edge [bend left] node[left] {} (n7)
	 edge [bend left] node[left] {} (n8)
	 edge [bend left] node[left] {} (n9)
	 edge [bend left] node[left] {} (n10)
	 edge [bend left] node[left] {} (n11)
	 edge [bend left] node[left] {} (n12)
	 edge [bend left] node[left] {} (n13);
\end{tikzpicture}
\end{center}
\subcaption{$G_2$}
\label{fig_med_3}
\end{minipage}

\caption{Figure \ref{fig_med_1} demonstrates the partition of the set of nodes into 6 sets. The first set (denoted $M$) contains nodes that will form the centers of the star networks chosen by the medium cost players $P_1$ and $P_2$. These star networks are depicted in Figures \ref{fig_med_2} and \ref{fig_med_3}. The networks of the remaining high-cost players have the same structure as the networks shown in Figures \ref{fig_struct_2} to \ref{fig_struct_5},  with $v_1$ and $v_2$ as isolated nodes.}
\label{fig5}
\end{figure}

\begin{example}
Consider a game with 13 nodes, 2 medium-cost players ($P_1$ and $P_2$) and 9 high-cost players ($P_3, \cdots, P_{11}$). Assume that the 9 high-cost players are the same as the high-cost players in Example \ref{ex:high_cost_game}. Based on Proposition \ref{prop:med_cost_player}, each of the medium-cost players $P_1$ and $P_2$ will have a star network centered on node $v_1$ and $v_2$, with peripheral nodes $V\setminus \{v_1\}$ and $V \setminus \{v_1, v_2\}$, respectively. These networks are shown in  Figure~\ref{fig_med_2} and \ref{fig_med_3}, respectively. The networks of the remaining players (which have high costs) have the same structure as in Example~\ref{ex:high_cost_game} with two extra isolated nodes, $v_1$ and $v_2$. 
Once again, we see that hub-and-spoke networks arise as a Nash equilibrium in this setting.  
\end{example}

The following corollary immediately follows from Propositions~\ref{prop:low_cost_player}, \ref{prop:high_cost_player} and \ref{prop:med_cost_player}.

\begin{corollary}
The multi-layer network formation game with strategic substitutes and distance-utilities has a pure Nash equilibrium for any set of players.
\label{cor:nash}
\end{corollary}

\section{Summary and Future Work}
\label{sec:conc}
In this paper, we introduced and studied the problem of strategic multi-layer network formation.  We generalized distance-based network formation to multi-layer networks, and showed that the problem of finding an optimal network in this setting is NP-hard. We characterized certain properties of optimal networks, and found the optimal networks for certain special cases of reference graphs.  Next, we formulated a multi-layer network formation game where each player builds a different layer of the network.  When the layers are viewed as strategic substitutes, we showed that the Nash equilibria of the game exhibit certain natural characteristics.  Specifically, the presence of low-cost players pushes high-cost players out of the game, and hub-and-spoke networks arise in the Nash equilibrium when there are no low-cost players.  

There are many interesting avenues for further research.  (1) Deriving approximation algorithms with provable performance guarantees is a natural approach to dealing with the inherent complexity of finding optimal networks; a deeper investigation of the connections between $t$-spanners and the best response network design problem might lead to such algorithms.  (2) Our initial simulations show that sequential best response dynamics converge to the Nash equilibria that we identified in this paper; providing formal proofs of convergence and understanding other non-simultaneous variants (such as Stackelberg games) is an important avenue for research.  (3)  While we have focused on distance-based utilities with strategic substitutes in this paper, it would also be of interest to study other classes of utility functions in the multi-layer network formation game.  (4) A mechanism to incorporate stochasticity and partial information into the network formation process would be of value in modeling and gaining further insights into the formation of realistic networks.

% or


\appendices  
\section{Proofs for Section~\ref{sec:properties}}
\label{app:proof_sectionV}

\subsection{Proof of Lemma \ref{lem2}}
\begin{IEEEproof}
Consider the utility of network $G_2$ with respect to $G_1$. Since there are no edges between the components in $G_{2}$, for any $(u,v) \in E_1$ with $u$ and $v$ in different components of $G_2$, $d_{G_2}(u,v)=\infty$. Thus $\sum_{(u,v) \in E_1} b(d_{G_2}(u,v))=\sum_{i=1}^k \sum_{(u,v) \in E_{1i}} b(d_{G_{2i}}(u,v))$, and the utility function can be written as
\begin{align*}
u(G_2|G_1) &= \sum_{(u,v) \in E_1} b(d_{G_2}(u,v)) - c|E_2|\\
& =\sum_{i=1}^k \left(\sum_{(u,v) \in E_{1i}} b(d_{G_{2i}}(u,v)) - c|E_{2i}|\right)\\
& = u(G_{21}|G_{11})+ \dots +u(G_{2k}|G_{1k}). \nonumber
\end{align*}
Now, if $G_{2i}$ is not a BR to $G_{1i}$ for some $i \in \{1,2, \dots, k\}$, replace it with a BR. This will increase the utility, contradicting the fact that $G_2$ is a BR.
\end{IEEEproof}

\subsection{Proof of Lemma \ref{lem4}}
\begin{IEEEproof}
Let $G_2=(N, E_2)$ be a BR network with respect to $G_1$, and suppose by way of contradiction that $v$ not isolated in $G_2$.  If $v$ is a leaf node in $G_2$ (i.e., it has a single neighbor), then the edge incident to $v$ is not used in any of the shortest paths between nodes in $N \setminus \{v\}$.  Removing that edge increases the utility of $G_2$ by $c$, contradicting the fact that it is a BR.

Now suppose that $v$ has two or more neighbors in $G_2$, and denote those neighbors by the set $J=\{v_{j_1}, v_{j_2}, \ldots, v_{j_l}\} \subseteq N \setminus \{v\}$ with $l \ge 2$.  
Construct a new network $G_3=(N, E_3)$ with
\begin{equation}
E_3 = (E_2 \setminus \{(v, v_{j_1}), (v,v_{j_2}), \ldots, (v,v_{j_l})\}) \cup \{(v_{j_1},v_{j_2}), (v_{j_1},v_{j_3}), \ldots, (v_{j_1},v_{j_l})\},
\end{equation}
i.e., we remove the $l$ edges from $v$ to its neighbors and add edges from $v_{j_1} \in J$ to the other nodes in $J$.  This results in a net removal of at least one edge from the graph.  Suppose that the shortest path between some pair of nodes in $N \setminus \{v\}$ passed through $v$ in $G_2$;  the shortest path now passes through $v_{j_1}$ in $G_3$, and is at least as short as the original shortest path.  Thus $u(G_3|G_1) > u(G_2|G_1)$ which contradicts the assumption that $G_2$ is a BR network to $G_1$. Therefore, $v$ must be an isolated node in $G_2$.
\end{IEEEproof}

\subsection{Proof of Lemma \ref{merged}}
\begin{IEEEproof}
In order to prove the first property, assume by way of contradiction that $G_2$ is a BR network and $G_2 \neq G_1$. Since $b(1)>b(2)$, by Lemma \ref{lem1}, we know that the number of edges in $G_2$ is less than in $G_1$. So there are vertices $u$ and $v$ such that $(u, v) \in E_1$ and $d_{G_2}(u,v) > 1$. Adding the edge $(u, v)$ to $E_2$ increases the utility by at least $b(1)-c- b(2)>0$ which contradicts the assumption that  $G_2 \neq G_1$ is a BR network. Therefore, the BR network must be equal to $G_1$.

For the second property note that if $G_2=G_1 \neq \phi$, then $u(G_2|G_1) = |E_1|(b(1)-c) < 0$ due to the assumption that $b(1)<c$. Thus it must be the case that $G_2 \neq G_1$, or $G_1$ is the empty network.

Finally in order to prove the third property, consider an arbitrary graph $G_1=(N, E_1)$ with $n$ nodes. By way of contradiction assume that $G_2 \neq \phi$ is a BR network with respect to $G_1$. Let $G_{21}=(N_1, E_{21})$ be a component of network $G_2$ with $1 < |N_1| \leq n$. By Lemma \ref{lem2}, we know that $G_{21}$ must be a BR to the subgraph induced by the node set $N_{1}$ on $G_1$, which we denote by $G_{11}=(N_{1}, E_{11})$. Thus
\begin{align} \nonumber
u(G_{21}|G_{11}) &\leq |E_{21}|(b(1)-c)+(|E_{11}|-|E_{21}|)b(2)\\\nonumber
&= |E_{21}|(b(1)-c+\alpha b(2))+(|E_{11}|-|E_{21}|(1+\alpha))b(2)\\\label{newalpha}
&= |E_{21}|(b(1)-c+\alpha b(2))+|E_{21}|\left(\frac{|E_{11}|}{|E_{21}|}-(1+\alpha)\right)b(2). 
\end{align}
Due to the assumption that $c>b(1)+\alpha b(2)$, the first term in \eqref{newalpha} is negative. Also, we have that
\begin{align*}
\frac{|E_{11}|}{|E_{21}|} \leq \frac{|E_{11}|}{|N_{1}|-1} = \frac{|E_{G_1}(N_{1}, N_{1})|}{|N_{1}|-1} \leq \operatorname*{max}_{2 \leq |S|, S \subseteq N}  \frac{|E_{G_1}(S, S)|}{|S|-1} =\alpha +1.
\end{align*}
The first inequality above follows from the fact that $G_{21}$ is a component and thus has at least $|N_{1}|-1$ edges. Thus the second term in equation~\eqref{newalpha} is nonpositive. Therefore, $u(G_{21}|G_{11}) <0$ which is a contradiction.  As a result $G_{21}$ (and thereby $G_2$) must be the empty network.
\end{IEEEproof}

\subsection{Proof of Lemma \ref{leafnode}}
\begin{IEEEproof}
Let the neighbor of $v$ in $G_1$ be denoted by $u$, and assume that network $H=(N, E_H)$ is a BR to network $G_1$.  We reason as we did in the proof of Lemma~\ref{lem4}, with a few additional details.

Consider the case that $b(1) \le c$.  Suppose that node $v$ is not isolated in $H$, and let $J = \{v_{j_1}, v_{j_2}, \ldots, v_{j_l}\} \subseteq N \setminus \{v\}$ be the neighbors of $v$ in $H$.  If $l = 1$ (i.e., $v$ has a single neighbor in $H$), the edge $(v, v_{j_1})$ is not used in any of the shortest paths between nodes in $N \setminus \{v\}$.  Removing that edge saves a cost of $c$, and loses at most a benefit of $b(1)$ (due to the loss of the path from $v$ to $u$ in $H$).  Since $b(1) \le c$, the resulting graph has utility at least as large as $H$.  

Now suppose $l > 1$.  Construct the new network $H_1=(N, E_{H_1})$ with edge set
\begin{equation}
E_{H_1} \triangleq \left(E_H \setminus\{(v,v_{j_1}), (v, v_{j_2}), \ldots, (v, v_{j_l})\}\right)
\cup \{(v_{j_1}, v_{j_2}), (v_{j_1}, v_{j_3}), \dots, (v_{j_1}, v_{j_l})\}.
\label{eqn:construct_H1}
\end{equation}
In other words, we remove all of the incident edges from $v$ in $H$ and add edges from each node in $J \setminus \{v_{j_1}\}$ to $v_{j_1}$.  This saves at least one edge, and $d_{H_1}(x,y) \leq d_{H}(x,y)$ for all $(x,y) \in E_1$.  
Thus, the only drop  in utility in graph $H_1$ arises from the loss of the path from node $v$ to $u$.  Again, since $b(1) \le c$, the graph $H_1$ has utility at least equal to the utility of the network $H$ and thus $H_1$ is also a best response.  The above two cases show that when $b(1) \le c$, there exists a best response where the leaf node $v$ is isolated.

Now consider the case where $b(1) \ge c$.  Then by Lemma \ref{connectedgraph} and Remark \ref{equalbc}, there exists a BR network $H=(N, E_H)$ containing a path from $v$ to $u$.  If $v$ is a leaf node in $H$, it is straightforward to show that there exists a BR network $H'$ where $v$ is connected to $u$.  Thus suppose $v$ is connected to the node set $J = \{v_{j_1}, v_{j_2}, \ldots, v_{j_l}\} \subseteq N \setminus \{v\}$ in $H$, with $l \ge 2$.  Construct a new graph $H_2 = (N, E_{H_2})$, where $E_{H_2} \triangleq E_{H_1} \cup \{(v,u)\}$ with $E_{H_1}$ as defined in \eqref{eqn:construct_H1}.  Arguing as above, the utility of $H_2$ is at least as high as the utility of $H$, and thus $H_2$ is a BR to $G_1$.  Since the edge $(v,u)$ cannot be in the shortest path between any pair of nodes in $N \setminus \{v\}$, we see that the subgraph of $H_2$ induced by $N \setminus \{v\}$ must be a best response to the corresponding subgraph of $G_1$.  This proves the result.
\end{IEEEproof}

\section{Proofs for Section~\ref{sec:charac}}
\label{app:proof_forest}

\subsection{Proof of Proposition \ref{G1forest}}

\begin{IEEEproof}
When $b(1)<c$, we use part 3 of Lemma~\ref{merged}. Following the same argument as in Example~\ref{ex:alpha} for trees, we have $\alpha = 0$ for $G_1$.  Thus the unique BR network to a forest is the empty network when $b(1)<c$.

For $b(1)-b(2)>c$, the unique best response to any network is the same network by the first part of Lemma~\ref{merged}. For $b(1)-b(2)\le c \leq b(1)$, note that $G_1$ is a 2-spanner forest of itself, and thus $G_1$ is a BR to itself by Lemma~\ref{lem:spanforestresult}, proving the second statement.  Since this BR has a utility of zero when $b(1) = c$, the empty network is also a BR for this value of $c$, proving the third statement.

Finally, we prove the uniqueness of the BR when $b(1) >\operatorname*{max} \{b(2), c\}$. If $G_1$ has $r$ connected components, then $|E_1|=|N|-r$. By Lemma \ref{connectedgraph}, we must have $|E_2| \geq |N|-r$. By Lemma \ref{lem1}, we know that $|E_2| \leq |E_1|=|N|-r$. Thus $|E_2|=|E_1|$ and since $b(1)>b(2)$, we  have $G_2=G_1$.
\end{IEEEproof}

\subsection{Proof of Proposition \ref{starmerged}}
\label{app:proof_star}

\begin{IEEEproof}
The first statement is a direct result of Lemma~\ref{merged}.

In order to prove the second statement we use Lemma~\ref{lem:spanforestresult}. Let $G^s$ be the star network centered at node $v$.  Since $G^s$ is a 2-spanner tree of $G_1$, it is a BR to $G_1$.

Next, we prove the third statement. Define $G^s$ as the star network centered at node $v$. By equation \eqref{P_2}, we have
\begin{equation}\label{uGsG1}
u(G^s|G_1)=(|N|-1)(b(1)-c)+(|E_1|-(|N|-1))b(2).
\end{equation}
Now assume that $G_2=(N, E_2)$ is a BR network. Using the same argument as in equation \eqref{eqlimit}, we have
\begin{align}
u(G_2|G_1) &\leq |E_2|(b(1)-c)+(|E_1|-|E_2|)b(2).  \label{uGG1}
\end{align}
Using equations \eqref{uGsG1} and \eqref{uGG1} we obtain
\begin{equation} \label{eqmin}
u(G^s|G_1)-u(G_2|G_1) \geq (|E_2|-(|N|-1))(b(2)-b(1)+c).
\end{equation}
According to the assumption of the Proposition, $c-b(1)\geq 0$ and thus the right hand side of equation \eqref{eqmin} is nonnegative for all $|E_2| \geq |N|-1$. Therefore, the utility of $G^s$ with respect to $G_1$ is as high as any other connected network.

Thus assume that $G_2$ is a non-empty disconnected network. Suppose that it has $\gamma$ components $G_{2k}=(N_k,E_{2k})$ for $k \in \{1,2,\cdots, \gamma\}$. Denote by $G_{1k}=(N_k, E_{1k})$, $k \in \{1,2,\cdots, \gamma\}$, the subgraphs induced by $N_k$ on $G_1$. Without loss of generality, let $v \in N_1$. Then, since $G_{11}$ contains a star subgraph (centered on $v$), and $G_{21}$ is a BR to $G_{11}$ (by Lemma~\ref{lem2}) and connected, we can take it to be a star by the above argument.  Next, we  aim to show that there exists a BR (constructed based on $G_2$) such that all of the components are isolated nodes except $G_{21}$. 

Suppose that some component of $G_2$ (not containing $v$) has more than one node and take this component to be $G_{22}$ without loss of generality. We know that $G_{22}$ is a BR to $G_{12}$ based on Lemma \ref{lem2}.  Arguing as in equation \eqref{eqlimit}, we have
\begin{equation}\label{kooft}
u(G_{22}|G_{12}) \leq |E_{22}|(b(1)-c)+(|E_{12}|-|E_{22}|)b(2).
\end{equation}
If $G_{22}$ has zero utility, we can replace it by the empty network and subsequently, we have the result. Thus assume by way of contradiction that it has some positive utility. Therefore, the right hand side of equation \eqref{kooft} is positive. Since $G_{22}$ is a connected network, $|E_{22}|\geq |N_2|-1$. Hence
\begin{align}
\frac{|E_{12}|-|E_{22}|}{|E_{22}|} \leq  \frac{|E_{12}|-(|N_2|-1)}{|N_2|-1} 
\leq  \frac{\binom{|N_2|}{2}-(|N_2|-1)}{|N_2|-1} < |N_2|-1. \label{newstarbound}
\end{align}
Using the assumption that the right hand side of inequality \eqref{kooft} is positive and by inequality \eqref{newstarbound}, we have that
\begin{align*}
0 &< |E_{22}|(b(1)-c)+(|E_{12}|-|E_{22}|)b(2)\\
&=|E_{22}|\left((b(1)-c)+\frac{|E_{12}|-|E_{22}|}{|E_{22}|} b(2) \right)\\
&< |E_{22}| \left(b(1)-c+(|N_2|-1)b(2)\right).
\end{align*}
Now consider a graph $\hat{G}_2$ obtained by removing all edges of $G_{22}$ and connecting all of its nodes to node $v$. Since $b(1)-c+(|N_2|-1)b(2)>0$ we have,
\begin{align} \nonumber
u(\hat{G}_2|G_1) &\geq \sum_{i \neq 2} u(G_{2i}|G_{1i})+|N_2|(b(1)-c)+|E_{12}|b(2)\\ \nonumber
&> \sum_{i \neq 2} u(G_{2i}|G_{1i})+|N_2|(b(1)-c)+|E_{12}|b(2)-(b(1)-c+(|N_2|-1)b(2))\\ \label{newchange2}
&= \sum_{i \neq 2} u(G_{2i}|G_{1i})+(|N_2|-1)(b(1)-c)+(|E_{12}|-(|N_2|-1))b(2), 
\end{align}
where the first inequality follows from the fact that the induced subgraph of $N_{1} \cup N_2$ on $\hat{G}_2$ is a connected network and we neglect the benefit (if any) from indirect connections between nodes in $N_1\setminus\{v\}$ and $N_2$. The second term in the first inequality captures the direct benefits and costs of the $|N_2|$ edges from nodes in $N_2$ to $v$, and the third term captures the benefits due to each pair of nodes in $N_2$ having a distance of $2$ from each other in $\hat{G}_2$ (via $v$). Next, note that 
\begin{align} \nonumber
u(G_{22}|G_{12}) &\leq |E_{22}|(b(1)-c)+(|E_{12}|-|E_{22}|)b(2)\\ \nonumber
&= (|N_2|-1)(b(1)-c)+(|E_{22}|-(|N_2|-1))(b(1)-c)+(|E_{12}|-|E_{22}|)b(2)\\ \nonumber
&\leq (|N_2|-1)(b(1)-c)+(|E_{22}|-(|N_2|-1))b(2)+(|E_{12}|-|E_{22}|)b(2)\\
&=(|N_2|-1)(b(1)-c)+(|E_{12}|-(|N_2|-1))b(2). \label{newchange3}
\end{align}
Substituting inequality \eqref{newchange3} in inequality \eqref{newchange2}, we have that
\begin{align*}
u(\hat{G}_2|G_1) &> \sum_{i \neq 2} u(G_{2i}|G_{1i})+(|N_2|-1)(b(1)-c)+(|E_{12}|-(|N_2|-1))b(2)\\
&\geq \sum_{i \neq 2} u(G_{2i}|G_{1i})+u(G_{22}|G_{12})=u(G_2|G_1).
\end{align*}
However this is a contradiction to the assumption that $G_2$ is a BR to $G_1$. Thus all of the nonempty components of $G_2$ (except $G_{21}$) must have zero utility and therefore, we can replace each of them by the empty network.
\end{IEEEproof}
\section{Proof of Proposition~\ref{prop:high_cost_player}}
\label{app:proof_high_cost}

To prove Proposition~\ref{prop:high_cost_player}, we will first need the following intermediate result.
\begin{lemma}\label{lem:lastlem}
Let $b(1)<c$. Consider network $G=(N, E)$  with components $G_i=(N_i, E_i)$ for $1 \leq i \leq r$ ($N=\cup_{i=1}^r N_i$ and $E=\cup_{i=1}^r E_i)$. Assume that every induced subgraph of $G$ has a 2-spanner forest. Then every BR of network $G$  is composed of a BR to each component of $G$. 
\end{lemma}
\begin{IEEEproof}
Let $F=(\cup_{i=1}^r N_i, E_F)$ be a BR to $G$. Suppose by way of contradiction that $F$ contains a non-empty component $F_1=(W, R)$ with nodes from $p$ different $N_i$ where $p \geq 2$. Let $G_{F_1}=(W, E_{F_1})$ denote the induced subgraph of $W$ on $G$ and $T$ be a 2-spanner forest of $G_{F_1}$. The spanner forest $T$ has $q$ components where $q \ge p$. Also note that $|R| \geq |W|-1 > |W|-q$. Then we have
\begin{align}\label{lastlem1}
u(F_1|G_{F_1}) &\leq |R|(b(1)-c)+(|E_{F_1}|-|R|)b(2)\\ \nonumber
&= (|R|-(|W|-q))(b(1)-c)+(|W|-q)(b(1)-c) +(|E_{F_1}|-|R|)b(2)\\ \nonumber 
&< (|R|-(|W|-q))b(2)+(|W|-q)(b(1)-c)+(|E_{F_1}|-|R|)b(2)\\ \nonumber 
&= (|W|-q)(b(1)-c)+(|E_{F_1}|-(|W|-q))b(2)\\\ \nonumber
&= u(T|G_{F_1}),
\end{align}
where the first inequality comes from the fact that at most $|R|$ pairs of nodes that are neighbors in $G_{F_1}$ have direct connections in $F_1$ and the remaining pairs of nodes are at a distance of at least 2 in $F_1$. The second inequality is due to $b(1)-c<b(2)$.

Inequality \eqref{lastlem1} means that by replacing $F_1$ with $T$, we can increase the utility of network $F$ which is a contradiction to the assumption that $F$ is a BR to $G$. Therefore, no component of $F$ contains nodes from multiple components in $G$ and thus by Lemma \ref{lem4}, the subgraph of $F$ induced by $N_i$ must be a BR to $G_i$ for $1 \le i \le r$, yielding the result.
\end{IEEEproof}

We are now in place to prove Proposition~\ref{prop:high_cost_player}.

\begin{IEEEproof}(Proposition~\ref{prop:high_cost_player})

Consider player $P_j$ where $1 \leq j \leq m$. If $j>i_1$ (i.e., $P_j \in H_0$), then  $G=\sim{\cup_{t=1, t \neq j}^m G_t}$ consists of disjoint complete graphs on node sets $V_r, V_{r-1}, \ldots, V_0$.  Since 
\begin{align*}
&k_j \geq k_{i_1+1}>n-i_1\\
&k_j \geq k_{i_2+1}>i_1-i_2\\
&\qquad\vdots\\
&k_j \geq k_{i_r+1}>i_{r-1}-i_r\\
&k_j \geq k_1 >i_r,
\end{align*}
a best response of player $P_j$ to any of these complete networks is the empty network (by Proposition~\ref{prop:jacksonresult} or Lemma~\ref{merged}). Every induced subgraph of $G$ has a star network on its non-empty components (which means it has a 2-spanner forest). Thus using Lemma \ref{lem:lastlem}, the empty network $G_j$ is a BR to the network of the other players.

Next, we prove that for player $P_j \in H_l$ where $1 \leq l \leq r-1$, the network $G_j$ is a BR to the other players' networks.  From the definition of the sets $H_l$ in \eqref{eqn:PV}, we have that $i_{l+1} < j \leq i_{l}$. Note that $G=\sim{\cup_{t=1, t \neq j}^m G_t}$, consists of disjoint complete graphs on node sets $V_{l+1}, \ldots, V_r$. It also has a component $C=(\cup_{t=0}^{l} V_t, E_{C})$ of size $n-i_{l+1}$. The structure of the network $C$ can be described as a set of complete networks of size $n-i_1+1, i_1-i_2+1,\ldots,i_{l-1}-i_{l}+1, i_{l}-i_{l+1}$ where all of them have the common node $v_j$. These complete networks are on node sets $V_0 \cup \{v_j\}, V_1\cup \{v_j\}, \ldots, V_{l-1}\cup \{v_j\}, V_{l}$.  Network $G$ satisfies the condition of Lemma~\ref{lem:lastlem} and thus a BR to $G$ can be obtained by finding a BR network to each component.
Since
\begin{align*}
&k_j \geq k_{i_{l+2}+1} >i_{l+1} -i_{l+2}\\
&\qquad\vdots\\
&k_j \geq k_{i_r+1}>i_{r-1}-i_r\\
&k_j \geq k_1>i_r,
\end{align*}
the best response of player $P_j$ to each of the complete networks on node sets $V_{l+1}, \ldots, V_r$ in $G$ is the empty network.   

Network $C$ has a star subgraph centered at node $v_j$ and hence by Proposition \ref{starmerged}, there exists a BR network $S=(\cup_{t=0}^{l} V_t, E_{S})$ that is a star network centered at node $v_j$ with potentially some isolated nodes.  Now assume that in the network $S$, there are edges from $v_j$ to a nonempty strict subset of nodes $R_q \subset V_q$ for some $0 \leq q \leq l$, and the set of nodes in $V_q \setminus R_q$ are isolated. Note that edges between node $v_j$ and the set of nodes $R_q$ are only useful for connections between nodes in $R_q \cup \{v_j\}$ and produces a utility of
\begin{equation}\label{final1}
|R_q| \left(b_j(1)-c_j+\frac{|R_q|-1}{2}b_j(2)\right) \ge 0, 
\end{equation}
where the inequality follows from the fact that this graph has utility at least as large as that of the empty network.  Now construct a new network $S'=(\cup_{t=0}^{l} V_t, E_{S'})$ by connecting a node $u \in V_q \setminus R_q$ to $v_j$, i.e., $E_{S'}=E_{S} \cup \{(v_j, u)\}$.  Then we have that $u(S'|C)-u(S|C)=b_j(1)-c_j+|R_q|b_j(2)$ which must be a positive value by inequality \eqref{final1}.  This contradicts the assumption that $S$ is a BR to $C$. Therefore, for each $0 \leq t \leq l$, node $v_j$ is either connected to all of the nodes in $V_t$  or to none of them. Since 
\begin{align*}
&k_j \leq k_ {i_1} \leq n-i_1+1\\
&k_j \leq k_{i_2} \leq i_1 -i_2+1\\
&\qquad\vdots\\
& k_j \leq k_{i_{l}} \leq i_{l-1}-i_{l}+1,
\end{align*}
a BR to all of the complete networks on nodes $V_t \cup \{v_j\}$ in $C$  is the star network for $0 \leq t \leq l-1$. However, since $k_j \geq k_{i_{l+1}+1} >i_{l} -i_{l+1}$, the BR to the complete network on the set of nodes $V_l$ is the empty network and thus all of the nodes in $V_l \setminus \{v_j\}$ must be isolated nodes.

Therefore, we can conclude that a star network centered on the node $v_j$ with peripheral nodes $\{v_{i_{l}+1},\ldots,v_n\}=\cup_{t=0}^{l-1} V_t$, and all other nodes being isolated is a BR to the network of the other players; this is precisely the network $G_j$ given in the statement of the proposition.   

Finally, we have to show that players $P_j$, $1 \leq j \leq i_r$ (i.e., $P_j \in H_r$) are in Nash equilibrium. Similar to the above, for player $P_j$,  $G=\sim{\cup_{t=1, t \neq j}^m G_t}$ consists of complete networks of size $n-i_1+1, i_1-i_2+1,\ldots,i_{r-1}-i_{r}+1, i_r$ with the common node $v_j$.  These complete networks are on node sets $V_0 \cup \{v_j\}, V_1\cup \{v_j\}, \ldots, V_{r-1}\cup \{v_j\}, V_{r}$. By an argument similar to the above, since $k_j \geq k_1 > i_r$ and 
\begin{align*}
&k_j \leq k_ {i_1} \leq n-i_1+1\\
&k_j \leq k_{i_2} \leq i_1 -i_2+1\\
&\qquad\vdots\\
& k_j \leq k_{i_r} \leq i_{r-1}-i_r+1,
\end{align*}
a star network centered on $v_j$ with peripheral nodes $\{v_{i_r+1},\ldots,v_n\}=\cup_{t=0}^{r-1} V_t$ (i.e., $G_j$) is a BR to the network of the other players.

Therefore, for each player $P_j \in P$, $G_j$ is a BR to $G=\sim{\cup_{t=1, t \neq j}^m G_t}$ and thus the given networks are in Nash equilibrium.
\end{IEEEproof}

\iffalse
\ifCLASSOPTIONcompsoc
  \section*{Acknowledgments}
\else
  \section*{Acknowledgment}
\fi

The authors would like to thank...

\ifCLASSOPTIONcaptionsoff
  \newpage
\fi
\fi



%\bibliographystyle{IEEEtran}

\bibliographystyle{IEEEtran}
\bibliography{refs}

\iffalse


\begin{IEEEbiography}{Michael Shell}
Biography text here.
\end{IEEEbiography}

\begin{IEEEbiographynophoto}{John Doe}
Biography text here.
\end{IEEEbiographynophoto}

%\newpage

\begin{IEEEbiographynophoto}{Jane Doe}
Biography text here.
\end{IEEEbiographynophoto}
\fi

%\vfill


\end{document}